\pdfoutput=1
\pdfoutput=1
\pdfoutput=1
\pdfoutput=1
\pdfoutput=1
\documentclass[twocolumn,10pt]{IEEEtran}
\normalsize

\usepackage{enumerate}
\usepackage{color}

\usepackage{cite}

\usepackage{float}
\newfloat{figtab}{htb}{fgtb}
\makeatletter
\newcommand\figcaption{\def\@captype{figure}\caption}
\newcommand\tabcaption{\def\@captype{table}\caption}
\makeatother

\usepackage{graphicx}
\usepackage{booktabs}
\usepackage{algorithm}
\usepackage{bm}
\ifCLASSINFOpdf
\else
\fi

\usepackage{amsthm}
\usepackage{amssymb,amsfonts}
\usepackage{amsfonts,balance}
\usepackage{subfigure}

\usepackage[cmex10]{amsmath}

\newtheorem{theorem}{Theorem}

\usepackage{algorithmic}

\usepackage{array}
\usepackage{multirow,makecell}
\usepackage[caption=false,font=footnotesize]{subfig}

\usepackage{verbatim}

\begin{document}
%
\title{Learning-based Prediction and Uplink Retransmission for Wireless Virtual Reality (VR) Network}
%
%

\author{Xiaonan~Liu, Xinyu~Li,~\IEEEmembership{Student Member,~IEEE,}
        Yansha~Deng,~\IEEEmembership{Member,~IEEE}\\
\thanks{X. Liu, X. Li and Y. Deng are with the Department of Engineering, King’s College London, London, WC2R 2LS, U.K. (e-mail:\{xiaonan.liu, xinyu.1.li, yansha.deng\}@kcl.ac.uk). (Corresponding author: Yansha Deng).}}

\maketitle

\begin{abstract}
Wireless Virtual Reality (VR) users are able to enjoy immersive experience from anywhere at anytime. However, providing full spherical VR video with high quality under limited VR interaction latency is challenging. If the viewpoint of the VR user can be predicted in advance, only the required viewpoint is needed to be rendered and delivered, which can reduce the VR interaction latency. Therefore, in this paper, we use offline and online learning algorithms to predict viewpoint of the VR user using real VR dataset. For the offline learning algorithm, the trained learning model is directly used to predict the viewpoint of VR users in continuous time slots. While for the online learning algorithm, based on the VR user's actual viewpoint delivered through uplink transmission, we compare it with the predicted viewpoint and update the parameters of the 
online learning algorithm to further improve the prediction accuracy. To guarantee the reliability of the uplink transmission, we integrate the Proactive retransmission scheme into our proposed online learning algorithm. Simulation results show that our proposed online learning algorithm for uplink wireless VR network with the proactive retransmission scheme only exhibits about 5$\%$ prediction error.
\end{abstract}



\begin{IEEEkeywords}
Viewpoint prediction, uplink retransmission, offline and online learning, virtual reality (VR).
\end{IEEEkeywords}

%
\IEEEpeerreviewmaketitle

\section{Introduction}
Since 2015, virtual reality (VR) has become increasingly popular, and the interactions between VR users and their world are being revolutionized with the development of VR technology \cite{VR2021,VRgrowth}. This vision has driven the commercial release of various VR hardware devices, including head-mounted displays (HMDs) such as HTC  Vive \cite{HTC} and Facebook Oculus Rift \cite{VR_device}. One of the main disadvantages of the wired HMDs is the constrained mobility of VR users, which severely affects the experience of VR users. To address this issue, wireless connected HMDs can be used to provide immersive experience from anywhere at anytime. However, one of the main challenges is to provide seamless and spherical VR video with high quality under limited VR interaction latency via fluctuated wireless channels \cite{Hu1}.

Meanwhile, when the VR user enjoys the VR video frame, it mainly focus on a certain direction at any given time slot. Based on the viewing direction, the corresponding portion of the image, defined by the Field of View (FoV) \cite{FOV}, needs to be rendered and delivered. The FoV determines the extent of the virtual environment that can be viewed. The center of the FoV that the VR user is watching is called Viewpoint \cite{Bao2}. If the viewpoint of the VR user is able to be well predicted, only its corresponding FoV part of the VR video is required to be rendered and delivered in advance, rather than rendering and transmitting the whole spherical video, which can further reduce the VR interaction latency \cite{XR}.

There are growing research interests in the wireless VR system. The authors in \cite{VR1} and \cite{VR2} proposed an echo state network (ESN) in wireless VR transmission to maximize the quality of service (QoS) and success transmission probability of VR users, respectively. While in \cite{VR6}, the joint caching and computing optimization problem of VR video frames was formulated to minimize the average required transmission rate to reduce communication bandwidth. Nevertheless, the authors in \cite{VR1,VR2,VR6} mainly focused on the resource allocation in the wireless VR system, and assumed that the tracking information, such as the VR users' head motion was sent to the small-cell base station (SBS) through uplink transmission without transmission error. 

The viewpoint prediction problem in wireless VR system has been studied in \cite{oculus-rift,LaViola,xiaonan,Bao1,Bao2,Bao3}. The authors in \cite{oculus-rift} considered viewpoint prediction via constant angular velocity and constant acceleration after every 20 $\rm{ms}$. In \cite{LaViola}, the authors proposed a double exponential smoothing method to predict users' head position and rotation after every 50 $\rm{ms}$. In \cite{xiaonan}, the authors considered Brownian Motion to simulate the eye movement and used Recurrent Neural Network (RNN) to predict viewpoint preference in continuous time slots. However, the methods in \cite{oculus-rift} and \cite{xiaonan} were not data-driven, and prediction results obtained in \cite{oculus-rift} and \cite{LaViola} would be unaligned with the viewpoint preference of VR users, which may not fit for real-time VR video transmission. In \cite{Bao2}, \cite{Bao1} and \cite{Bao3}, the authors only used offline Linear Regression (LR) \cite{LR} and Neural Network (NN) \cite{NN} to predict the viewpoint of VR users in continuous time slots with real VR dataset, and assumed that all the viewpoint requests are available at the SBS, which is not possible without $100\%$ reliable uplink transmission. Based on such predictions in \cite{Bao2}, \cite{Bao1} and \cite{Bao3}, the authors minimized the multicast bandwidth consumption by sending the predicted part of the spherical VR video.

Because of the random nature of the head motion of VR users, viewpoint prediction based on delayed uplink viewpoint transmission may be prone to error, and only using trained LR and NN cannot guarantee the highest prediction accuracy and capture the complex dynamics viewpoint preference over time, which may further degrade the quality of experience (QoE) \cite{Hu1} of VR users. To address this issue, a RNN based on the state-of-the-art Long Short-Term Memory (LSTM) \cite{LSTM} or Gated Recurrent Units (GRU) \cite{GRU} architecture can be designed to predict the viewpoint of the VR user. However, the trained LR, NN, LSTM and GRU learning models, namely, offline learning models, cannot adapt to the dynamic changing environment, and has poor adaptibility to viewpoint prediction of new VR users. 

In contrast to the offline learning algorithm, online learning model is updated with each FoV request received, and can automatically adjust the model itself according to the change of the received data. In \cite{xiaonan}, through transmitting the FoV request to the mobile edge computing (MEC)-enabled SBS in the wireless VR network, the MEC was able to accurately predict the required FoV of the VR user over time, render and deliver the FoV in advance, which could decrease the VR interaction latency and improve the QoE of the VR user. Therefore, the online learning algorithm has the potential to learn and update the best predictor for future FoV preference at each time slot, and can be updated instantly once the FoV request of new VR user is received \cite{online1,online2,online3}.

To update the hyper-parameters in an online fashion, the VR users need to transmit its actual viewpoint to the SBS through uplink transmission, and the SBS will compare the actual viewpoint with the predicted viewpoint to reduce the loss between them. In the wireless VR network, due to the unstable wireless channels and the interference from other VR users, it is possible that the uplink transmission between the VR user and the SBS fails. Nevertheless, in the aforementioned wireless VR systems \cite{oculus-rift,VR1,VR2,VR6,LaViola,xiaonan,Bao1,Bao2,Bao3}, the authors did not consider the potential uplink transmission failure caused by the wireless fluctuation to the online training, and the potential uplink transmission enhancement for better online training. To deal with this issue, the Proactive retransmission scheme \cite{URLLC1,URLLC2,URLLC3,URLLC4} is first proposed for the uplink viewpoint transmission to achieve ultra-reliable low-latency communication (URLLC) requirement for this type of small data transmission.

Motivated by above, in this paper, we develop offline and online learning algorithms for a wireless uplink VR system under proactive retransmission scheme to efficiently maximize the viewpoint prediction accuracy of VR users. The main contributions can be summarized as follows:
\begin{itemize}
    \item Based on the historical and current viewpoint of the VR user in the real VR dataset, we develop offline and online learning algorithms to predict the viewpoint of the VR user in continuous time slots, in order to capture the dynamical viewpoint preference of VR users over time.
    \item  There are 16 VR videos in the real VR dataset, each VR video has its own property, and we first learn separate learning model for each VR video to predict the viewpoint of its corresponding VR users. When the number of VR videos increases, one learning model for each VR video may occupy much more computation resource and memory of the SBS. Therefore, to evaluate the generality of the viewpoint prediction model, we further propose one learning model for all VR videos. To ensure that the FoV request of each VR user in the VR dataset has the opportunity to be tested and avoid any biased performance, we use K Cross Validation to train the learning model.
    \item According to \cite{Bao2}, the viewpoint of each VR user has strong short-term auto-correlation, which means that the viewpoint can be well predicted based on the historical viewpoint of each VR user. For the offline learning algorithms, we train the $n$-order Linear Regression (LR), Neural Network (NN), and Recurrent Neural Network (RNN) based on the state-of-the-art Long-short Term Memory (LSTM)/Gated Recurrent Unit (GRU) architecture to predict the viewpoint of the VR user over time. However, the offline learning model cannot adapt to the dynamic changing environment when new VR users exist.
    \item In the online learning algorithms, we take into account the effect of the failure during the uplink transmission on viewpoint prediction. The uplink transmission may fail because of the unstable wireless channels and interference, which can result in incomplete training data, and may decrease the prediction accuracy of the learning model. To guarantee the reliability of the uplink transmission, we introduce the proactive retransmission scheme to the uplink VR viewpoint transmission during the online learning. Interestingly, our results shown that the online GRU algorithm for uplink wireless VR network with the proposed retransmission scheme can achieve 95$\%$ prediction accuracy.
\end{itemize}

The rest of this paper is organized as follows. The VR data description and analysis are proposed in Section II. The system model and problem formulation are presented in Section III. Learning algorithms for viewpoint prediction is proposed in Section IV. The simulation results and conclusions are described in Section V and Section VI, respectively.

\begin{figure}[h]
    \centering
    \includegraphics[width=3.0 in]{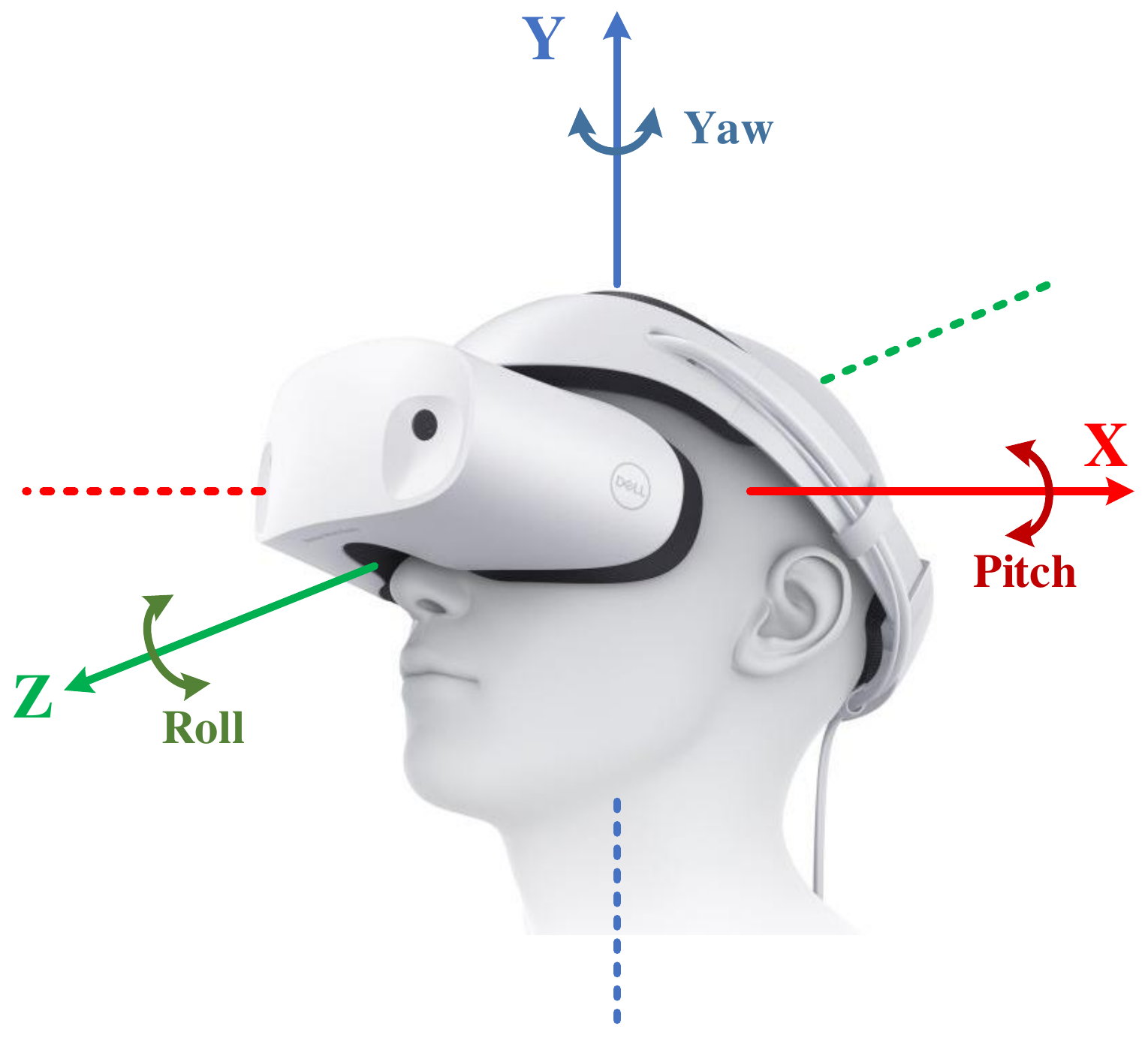}
    \caption{VR user viewing direction.}
    \label{basic_modules}
\end{figure}

\section{VR Data Description and Analysis}

\begin{figure*}[h]
    \centering
    \includegraphics[width=4.5 in]{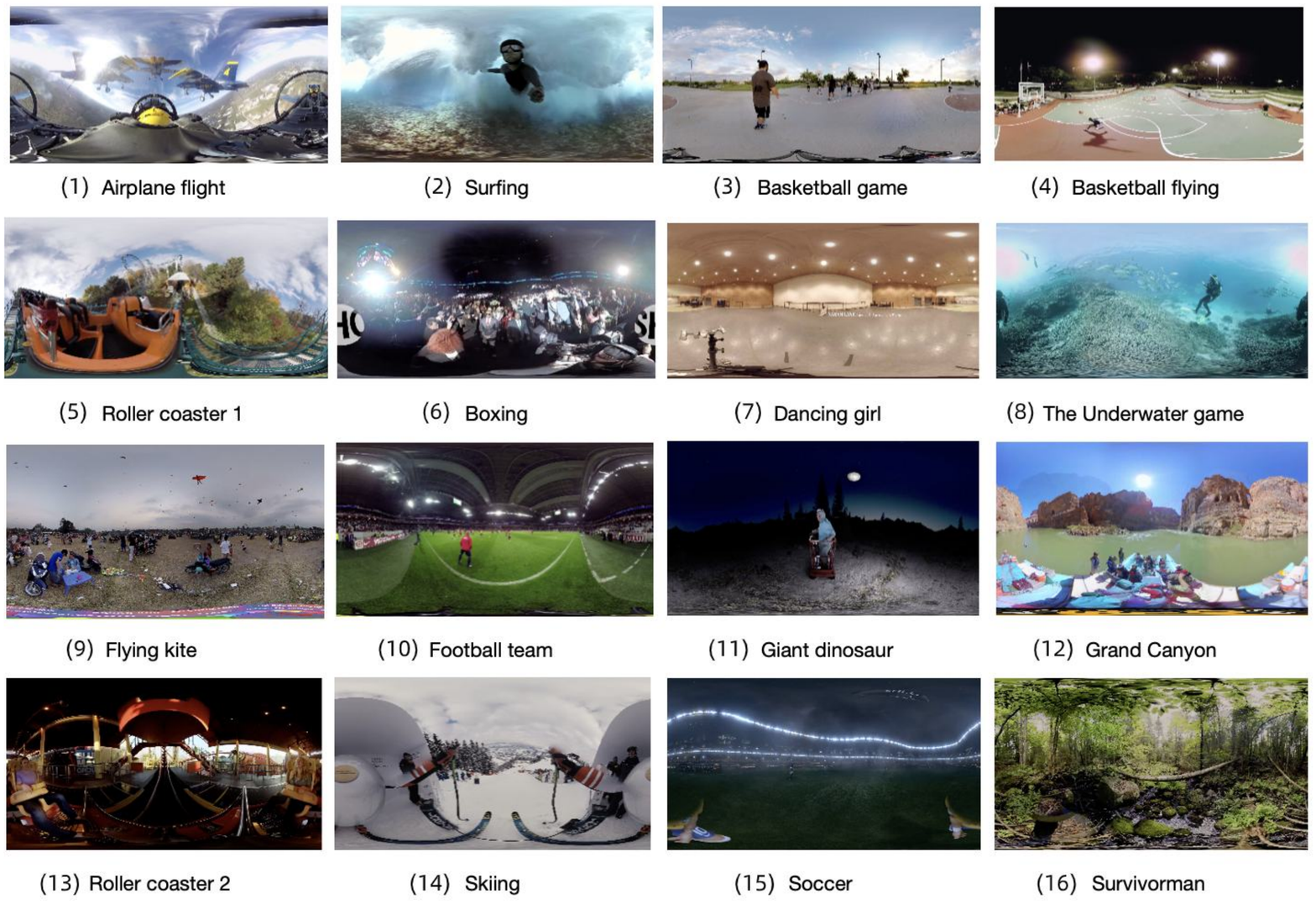}
    \caption{VR video screenshot.}
    \label{basic_modules}
\end{figure*}

The VR dataset obtained from \cite{VRdataset} includes $16$ clips of VR videos with $153$ VR users, and 969 data samples of the motion in three dimensions, pitch, yaw, and roll, namely, $X$, $Y$ and $Z$ viewing angles, which are shown in Fig. 1. Each dimension is presented by an angle ($-180^{\circ}$ to $180^{\circ}$), and each data includes the $X$, $Y$ and $Z$ viewing angles of each VR user at each time slot.

\subsection{VR Video Description}
The scene of the VR video is shown in Fig. 2. From Fig. 2, we can observe that the VR videos can be divided into three categories: 1) 7 of them are sports content, including Surfing, Basketball, Boxing, Football, Skiing, and Soccer; 2) 2 of them are Landscapes content, including Grand Canyon, and Survivorman; and 3) 5 of them are Entertainment, including Airplane flight, Underwater game, Roller coaster, Dancing girl, Flying Kite, and Glant Dinosaur.

\begin{table}
\centering
\caption{VR Video Information}
\begin{tabular}[c]{c|c|c|c}
\hline
\hline Video Number & VR Video Scene & Resolution & Bitrate \\
\hline (1) & Airplane Flight & 4k & 13.04~\rm{Mbps} \\
\hline (2) & Surfing & 4k & 23.2~\rm{Mbps} \\
\hline (3) & Basketball Game & 4k & 9.2~\rm{Mbps} \\
\hline (4)& Basketball Flying & 4k & 5.42~\rm{Mbps} \\
\hline (5)& Roller Coaster1 & 4k & 31.85~\rm{Mbps} \\
\hline (6)& Boxing & 4k & 4.4~\rm{Mbps} \\
\hline (7)& Dancing Girl & 4k & 6.58~\rm{Mbps} \\
\hline (8)& The Underwater World & 4k & 23.47~\rm{Mbps} \\
\hline (9)& Flying Kite & 4k & 10.98~\rm{Mbps} \\
\hline (10)& Football Team & 4k & 8.24~\rm{Mbps}\\
\hline (11)& Giant Dinosaur & 1080p & 1.35~\rm{Mbps}\\
\hline (12)& Grand Canyon & 2k & 5.12~\rm{Mbps}\\
\hline (13)& Roller Coaster2 & 4k & 22.87~\rm{Mbps}\\
\hline (14)& Skiing & 4k & 21.08~\rm{Mbps}\\
\hline (15)& Soccer & 4k & 12.76~\rm{Mbps}\\
\hline (16)& Survivorman & 4k & 29.1~\rm{Mbps}\\
\hline
\hline
\end{tabular}
\end{table}

The 16 VR video clips are downloaded from YouTube. The duration of each VR video is 30 seconds and each VR video is divided into 300 equal parts, which means that each sample point of the VR video lasts for 0.1 second. Among these VR videos, 14 of them are 4$\rm{K}$ resolution, one of them is 2$\rm{K}$ resolution, and one of them is 1080$\rm{P}.$ The detailed attributes of each VR video are shown in Table I.

In the experimental data, 153 VR users watched these VR videos, where 35 of them enjoyed all 16 VR video clips, and 118 of them enjoyed 3 to 5 randomly selected VR video clips. The detailed number of VR users watching each VR video is shown in Fig. 3. We can obtain that each VR video is watched by an average of 60 VR users, with a minimum of 46, and a maximum of 84. Meanwhile, the age distribution of all VR users is shown in Fig. 4. From Fig. 4, we can see that 
more than half of VR users are between 20 and 30. In addition, for all the VR users, $38\%$ VR users are female, and $34\%$ VR users wear glasses.

\begin{figure}[!t]
    \centering
    \includegraphics[width=3.5 in]{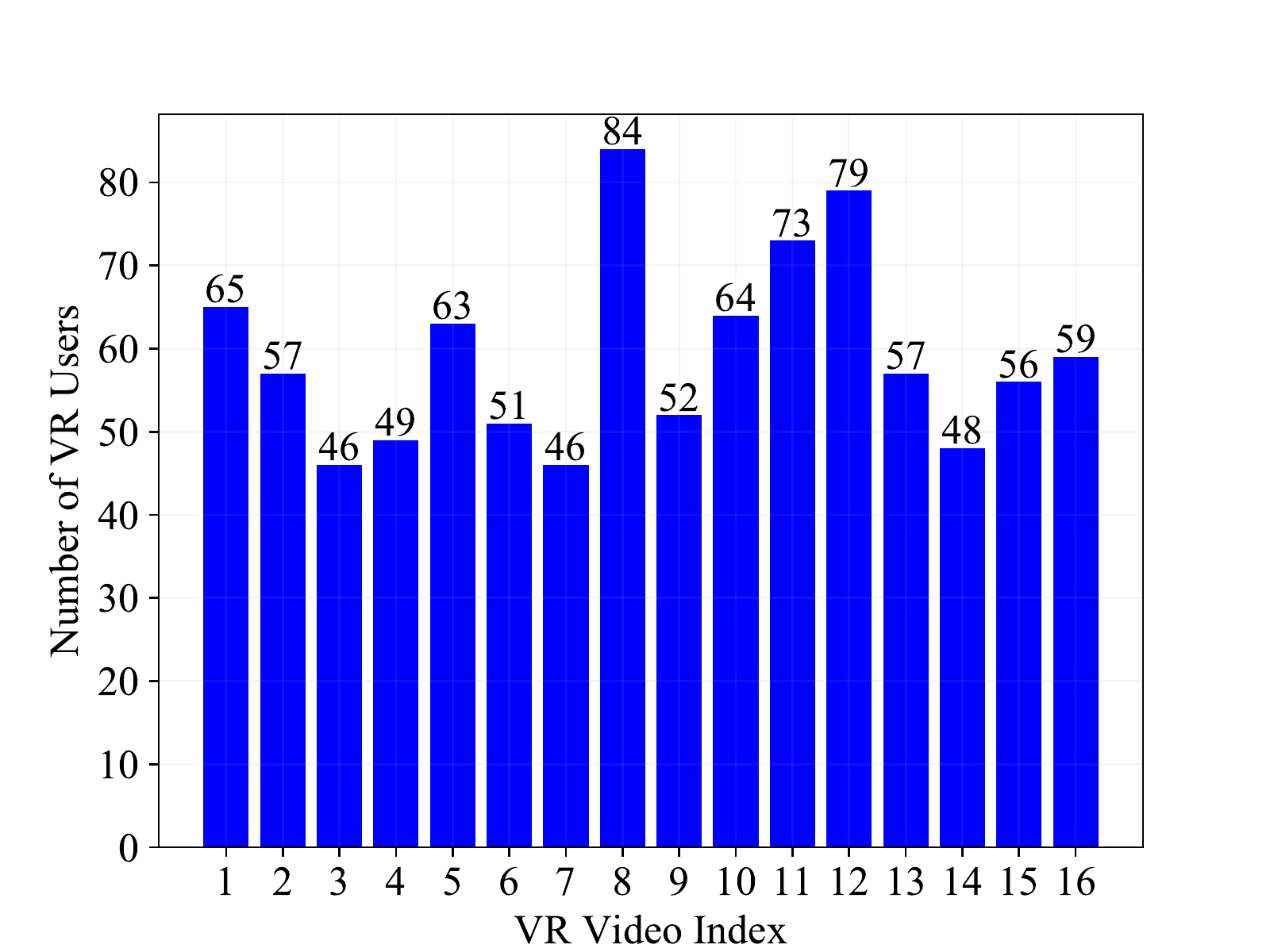}
    \caption{The detailed number of VR users watching each VR video.}
    \label{basic_modules}
\end{figure}

\begin{figure}[!t]
    \centering
    \includegraphics[width=3.5 in]{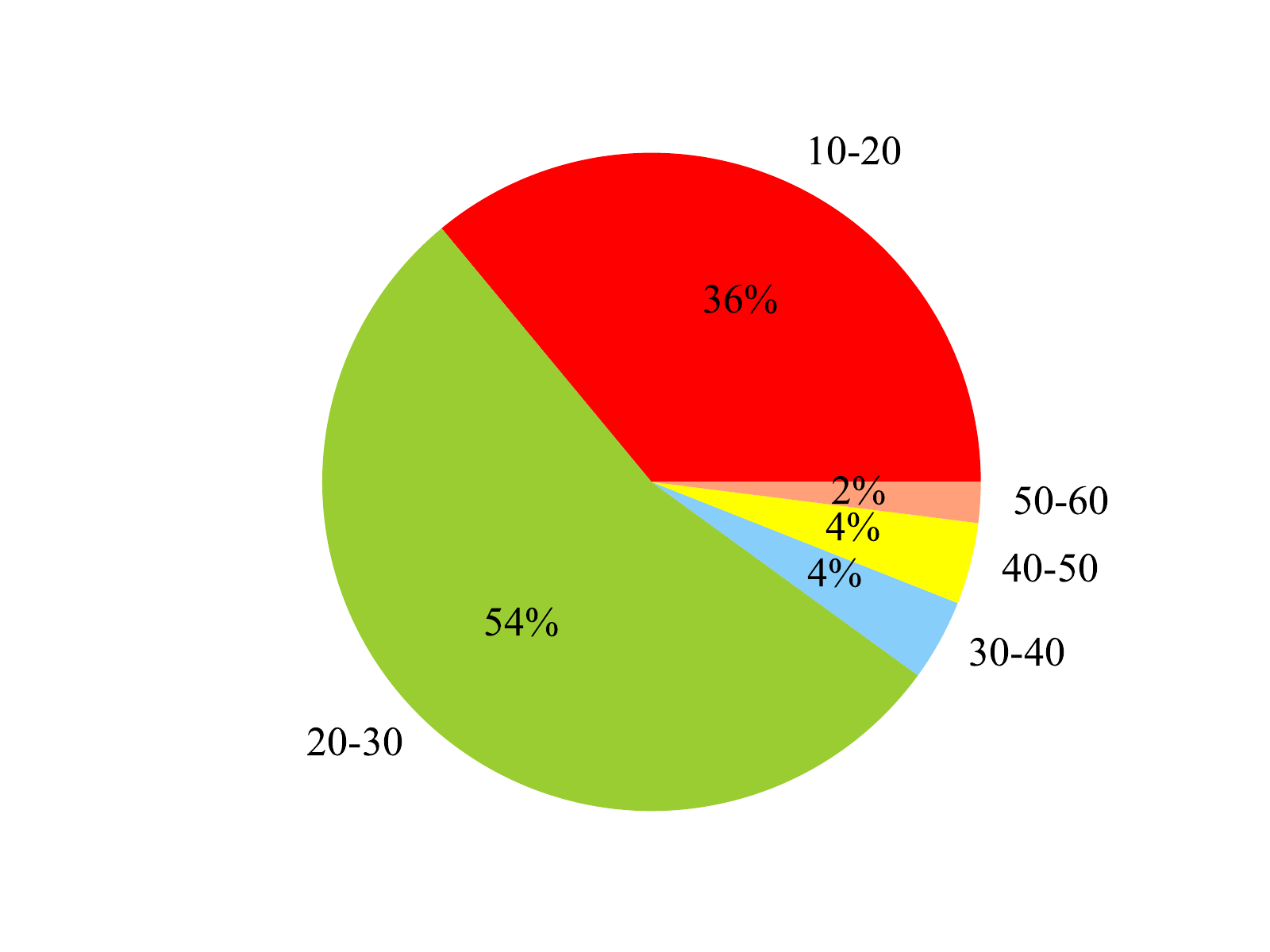}
    \caption{Age distribution of all VR users.}
    \label{basic_modules}
\end{figure}

\subsection{Viewpoint Distribution}
In the VR dataset, most VR users have similar viewpoint when enjoying the same video. We plot the viewpoint of all VR users for the 16 VR videos in $X$, $Y$ and $Z$ angles, respectively, which are shown in Fig. 5, 6 and 7, respectively. The $\rm{X}$-axis and $\rm{Y}$-axis of the viewpoint distribution figure are VR video playout time and degree of angles at each time slot, respectively. From Fig. 5, 6 and 7, we can know that the viewpoint range of $X$, $Y$ and $Z$ angles are (-$50^{\circ}$, $50^{\circ}$), (-$150^{\circ}$, $150^{\circ}$) and (-$50^{\circ}$, $50^{\circ}$), respectively.

\begin{figure}[!t]
	\centering
	\subfigure{\includegraphics[width=0.83in]{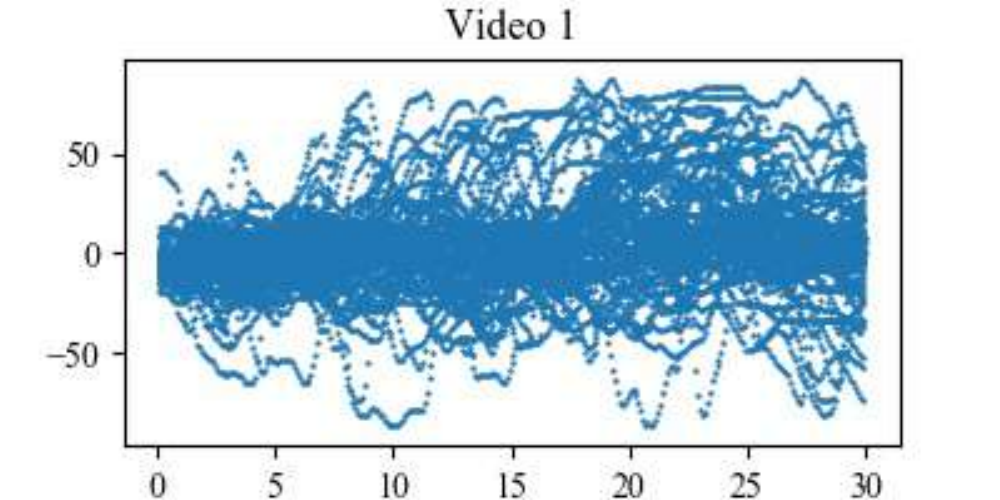}}\label{fig_first_case}
	\subfigure{\includegraphics[width=0.83in]{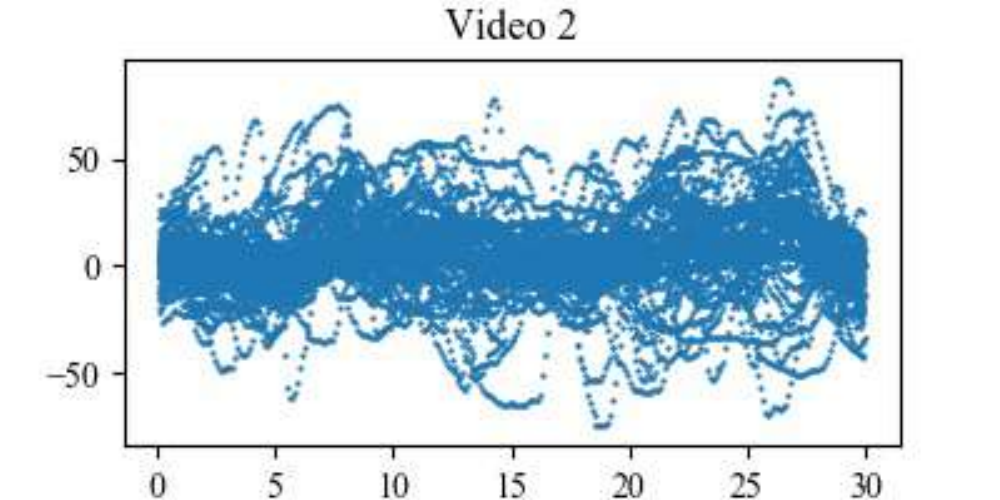}}\label{fig_second_case}
	\subfigure{\includegraphics[width=0.83in]{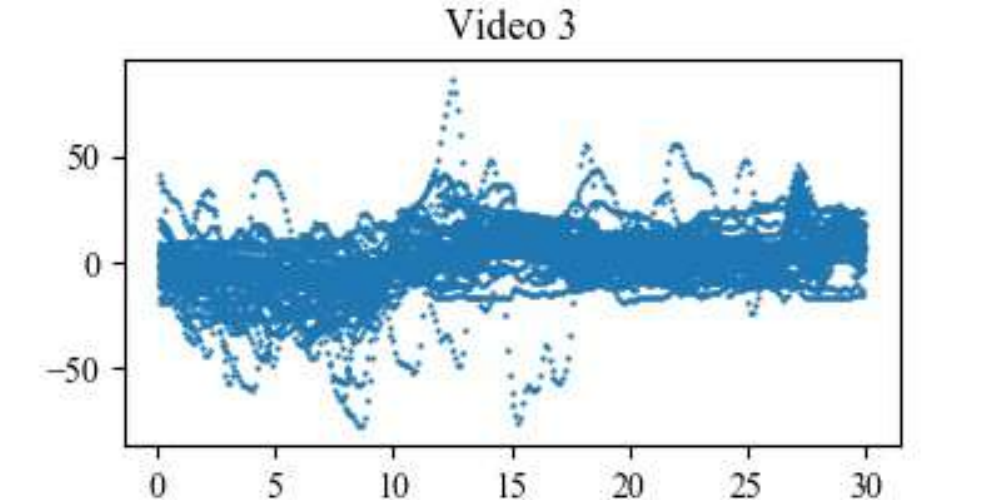}}\label{fig_third_case}
	\subfigure{\includegraphics[width=0.83in]{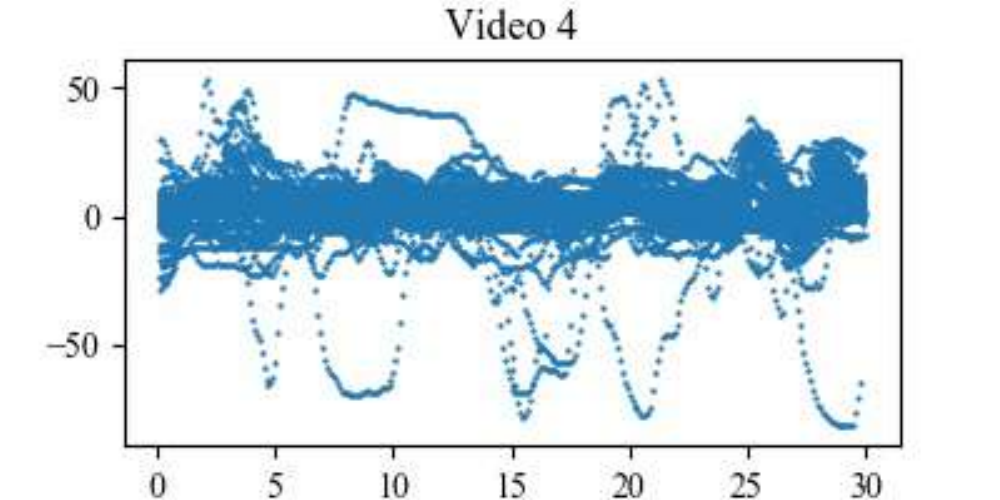}}\label{fig_third_case}
	
	\subfigure{\includegraphics[width=0.83in]{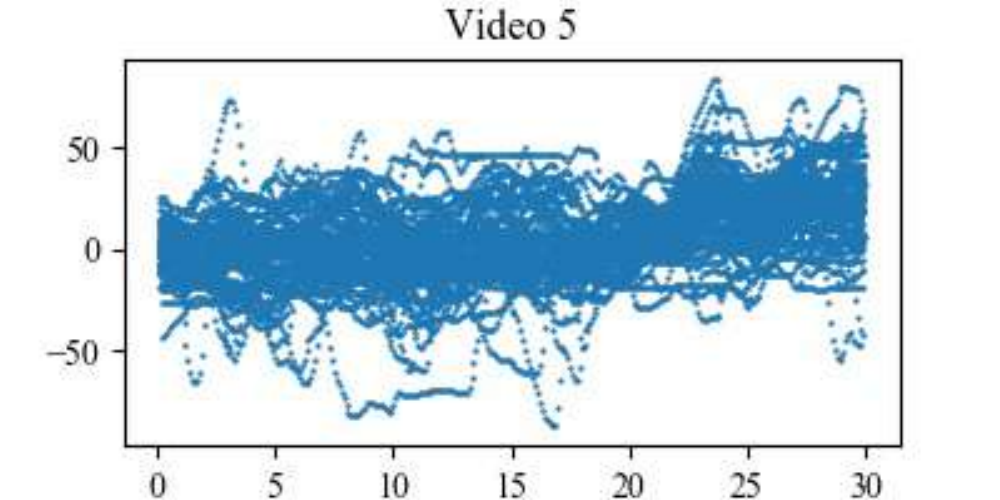}}\label{fig_first_case}
	\subfigure{\includegraphics[width=0.83in]{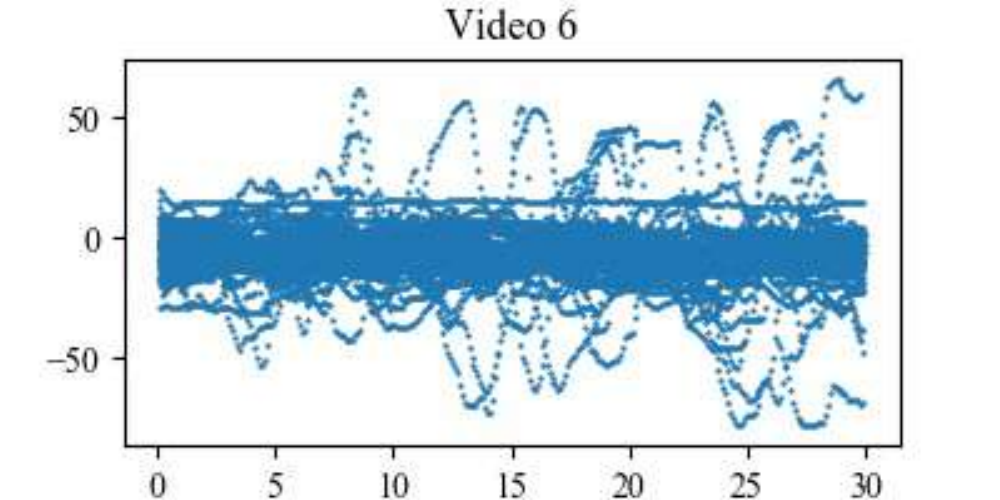}}\label{fig_second_case}
	\subfigure{\includegraphics[width=0.83in]{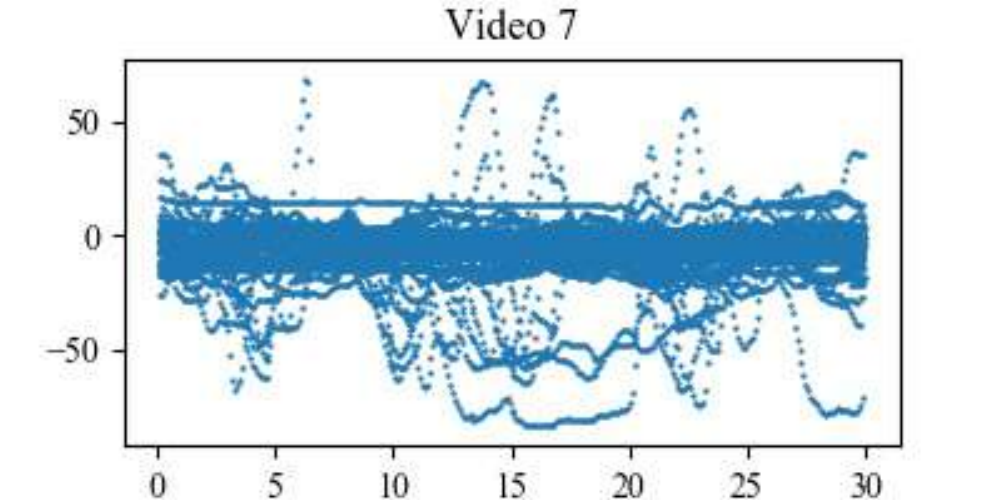}}\label{fig_third_case}
	\subfigure{\includegraphics[width=0.83in]{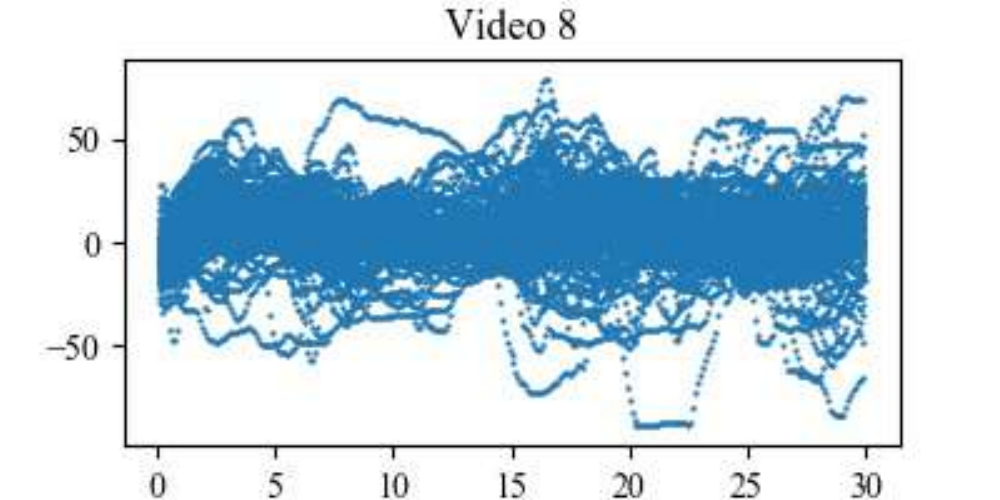}}\label{fig_third_case}
	
	\subfigure{\includegraphics[width=0.83in]{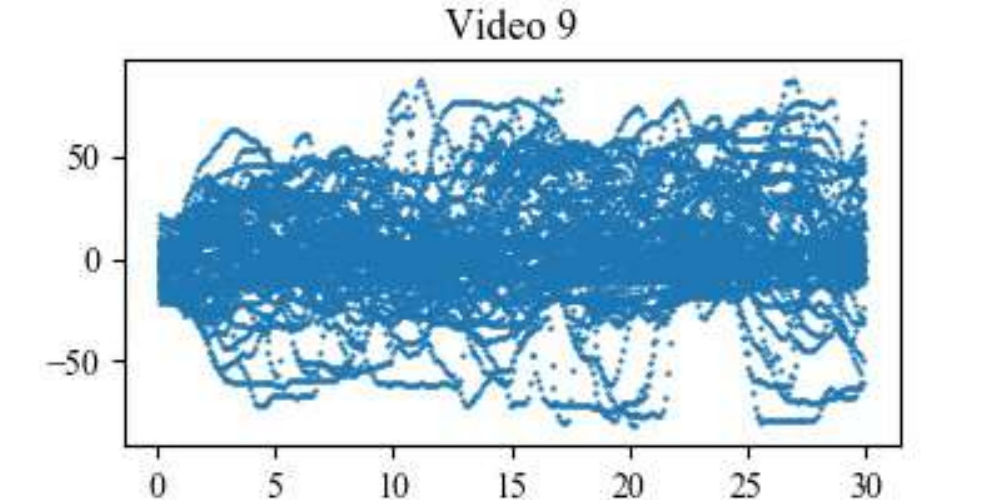}}\label{fig_first_case}
	\subfigure{\includegraphics[width=0.83in]{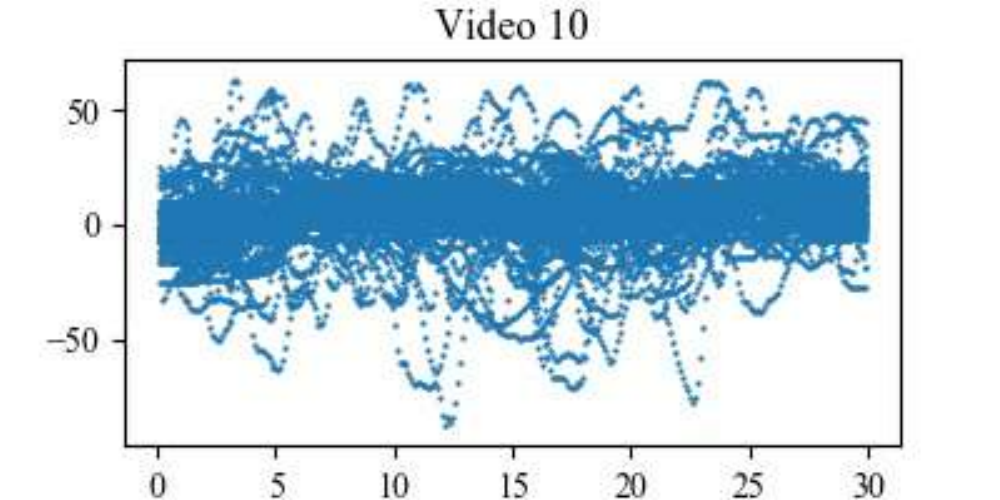}}\label{fig_second_case}
	\subfigure{\includegraphics[width=0.83in]{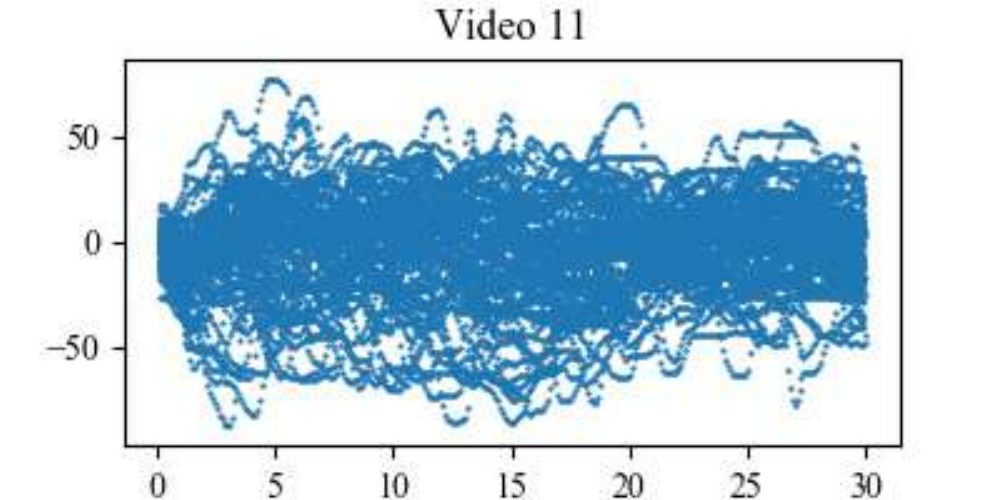}}\label{fig_third_case}
	\subfigure{\includegraphics[width=0.83in]{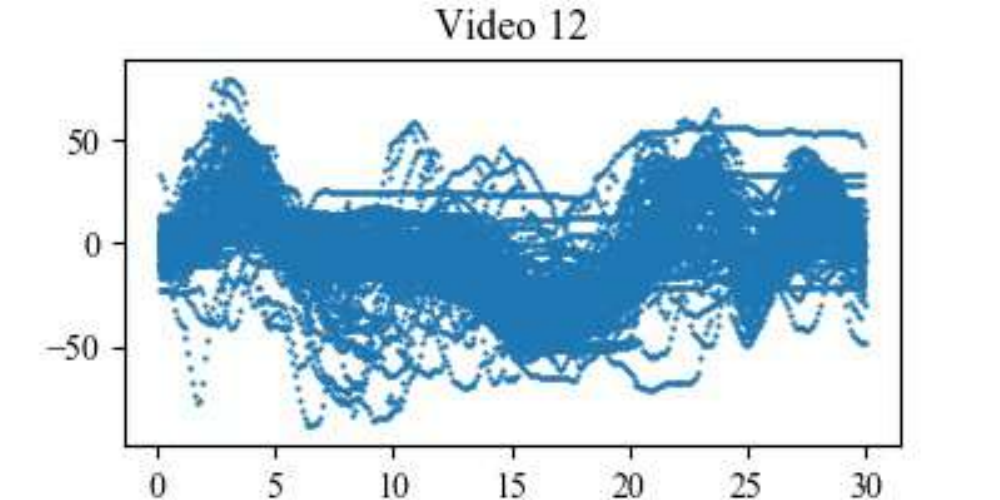}}\label{fig_third_case}
	
	\subfigure{\includegraphics[width=0.83in]{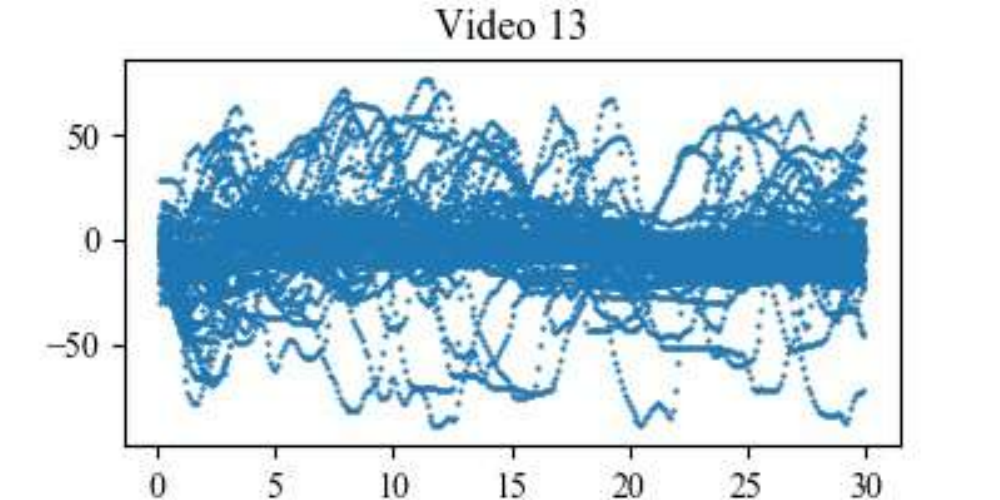}}\label{fig_first_case}
	\subfigure{\includegraphics[width=0.83in]{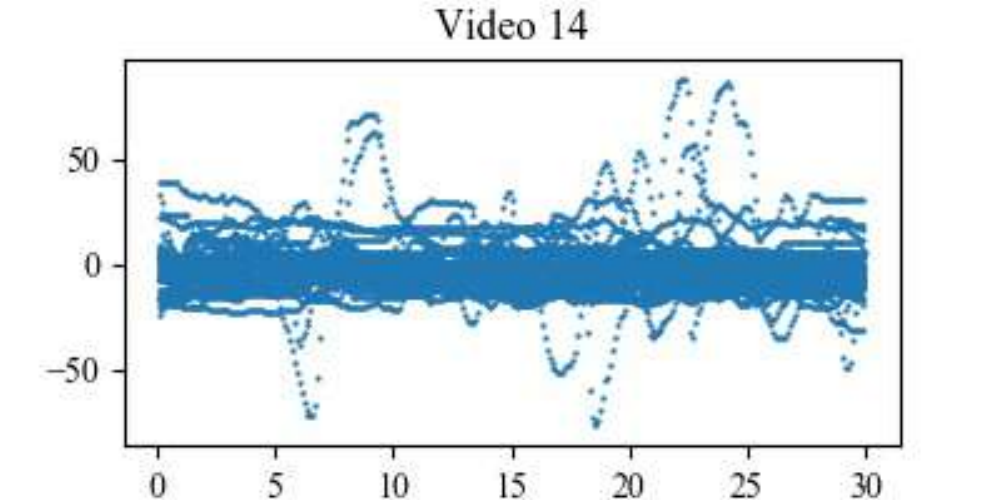}}\label{fig_second_case}
	\subfigure{\includegraphics[width=0.83in]{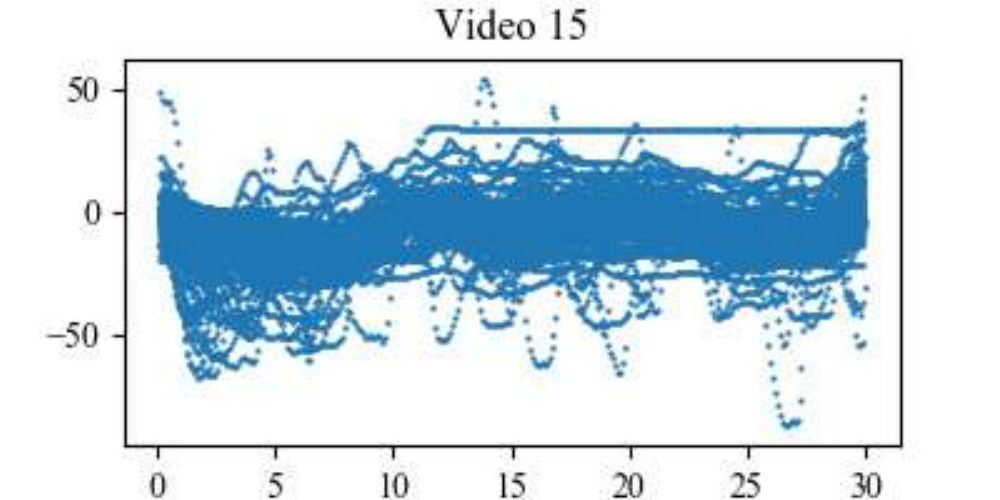}}\label{fig_third_case}
	\subfigure{\includegraphics[width=0.83in]{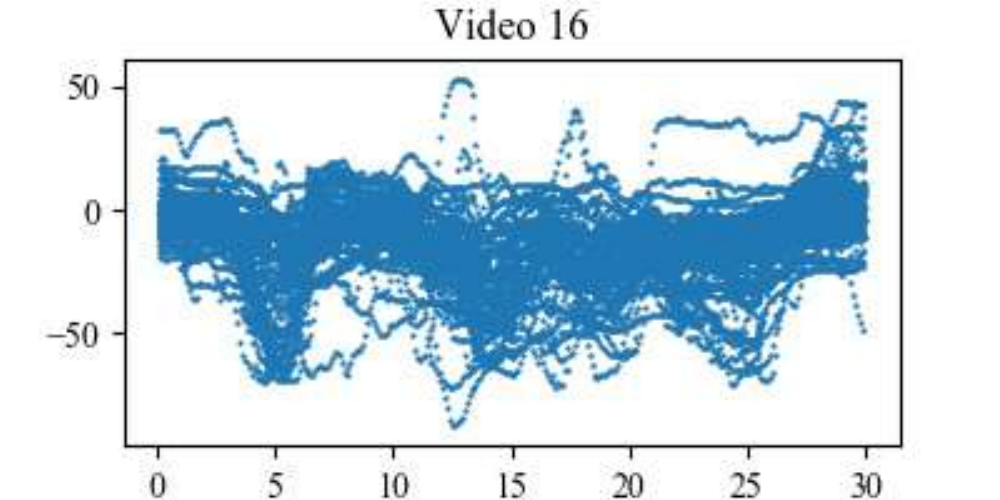}}\label{fig_third_case}
	\caption{$X$ angle distribution of all VR users.}
	\label{basic_modules}
\end{figure}

\begin{figure}[!t]
	\centering
	\subfigure{\includegraphics[width=0.83in]{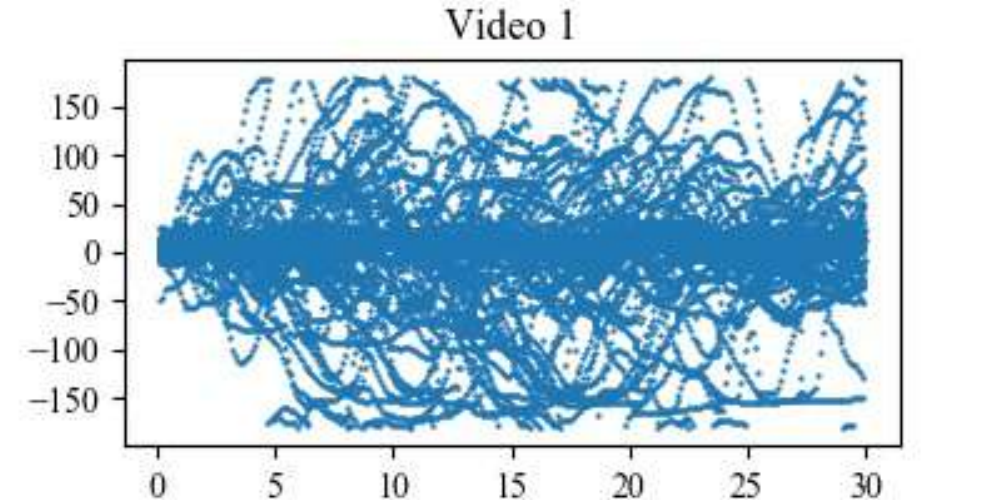}}\label{fig_first_case}
	\subfigure{\includegraphics[width=0.83in]{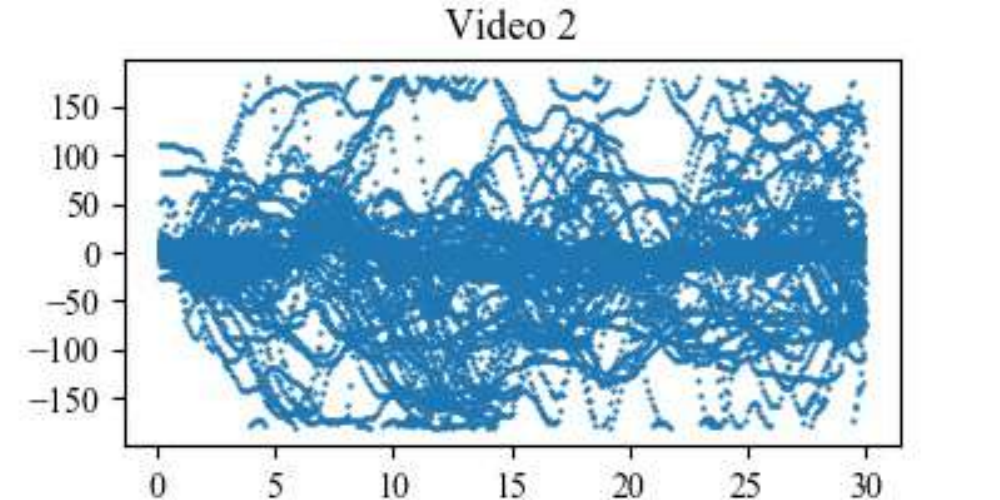}}\label{fig_second_case}
	\subfigure{\includegraphics[width=0.83in]{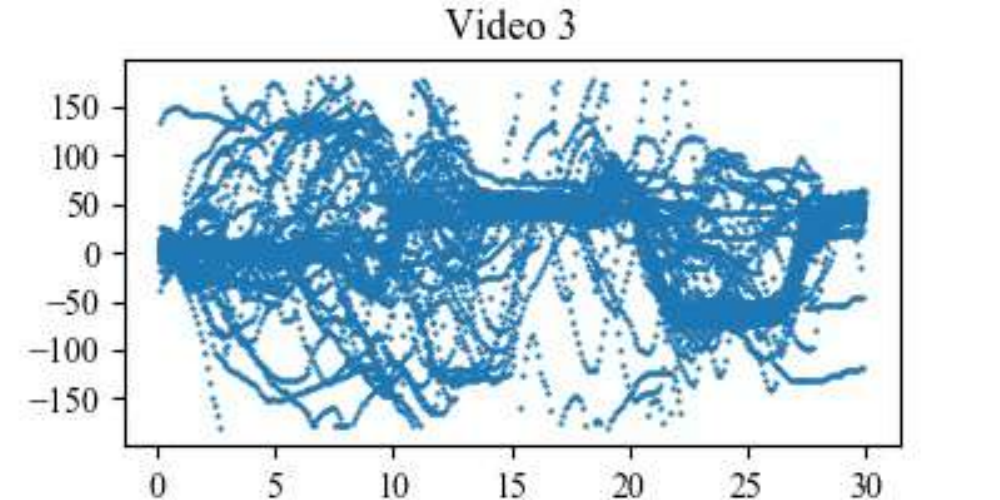}}\label{fig_third_case}
	\subfigure{\includegraphics[width=0.83in]{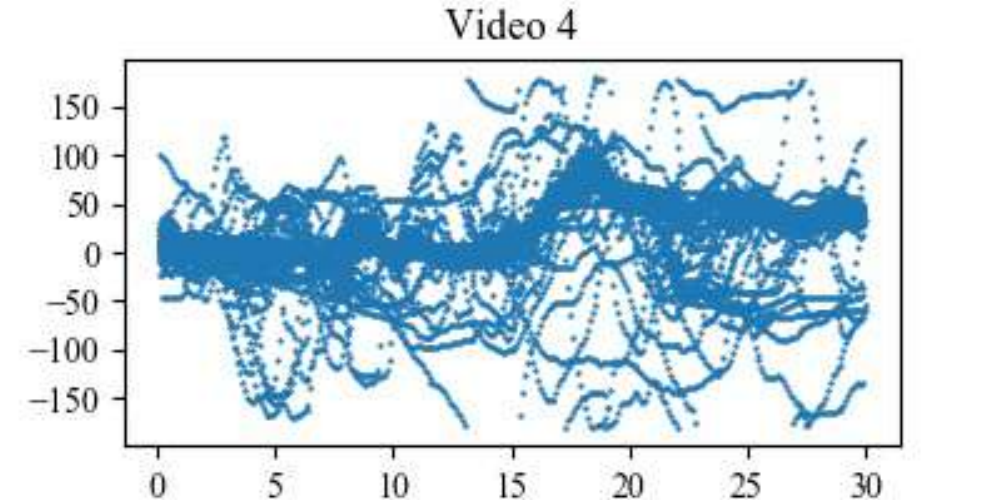}}\label{fig_third_case}
	
	\subfigure{\includegraphics[width=0.83in]{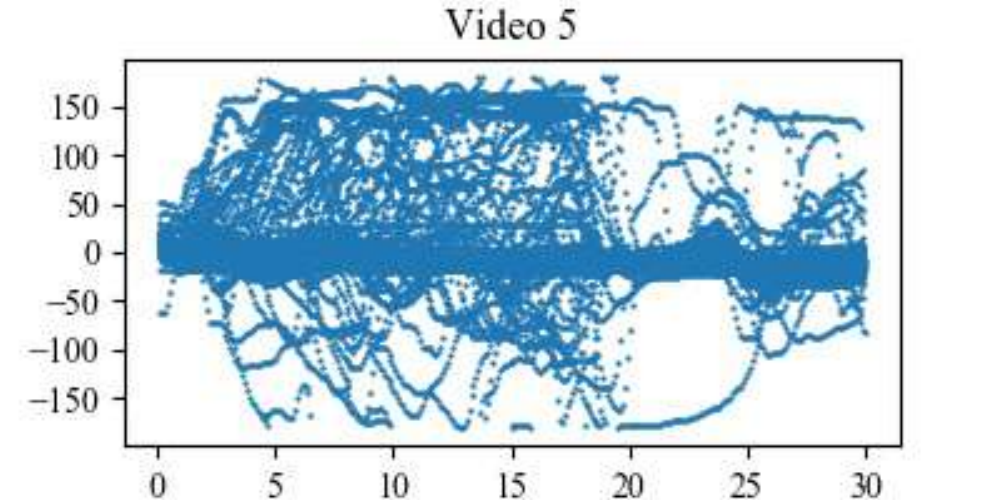}}\label{fig_first_case}
	\subfigure{\includegraphics[width=0.83in]{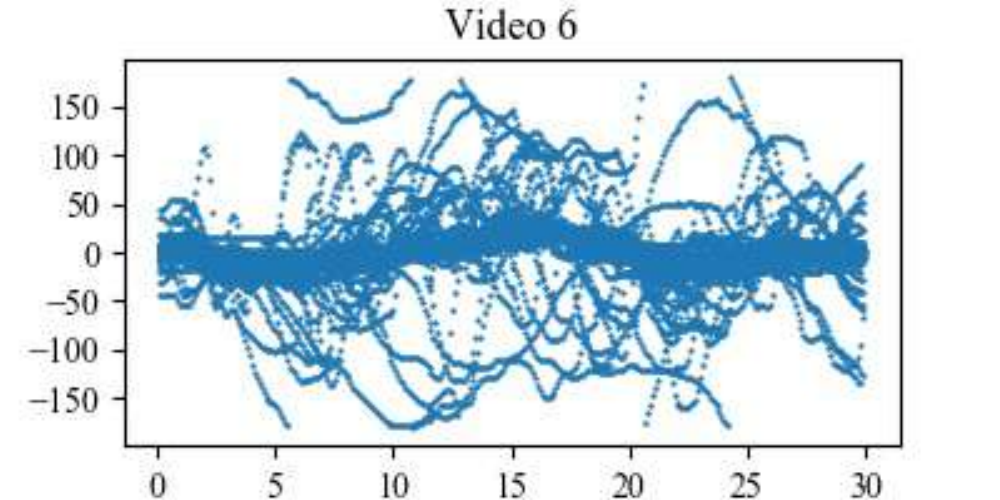}}\label{fig_second_case}
	\subfigure{\includegraphics[width=0.83in]{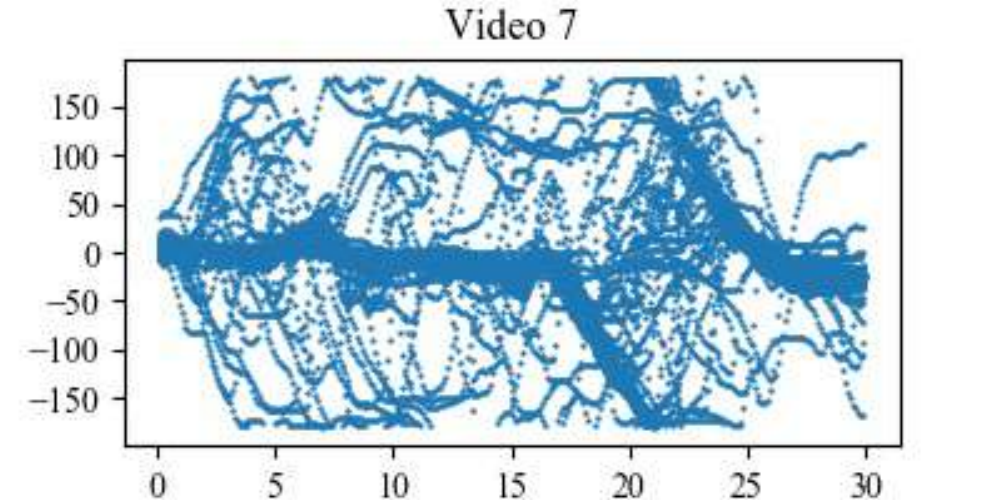}}\label{fig_third_case}
	\subfigure{\includegraphics[width=0.83in]{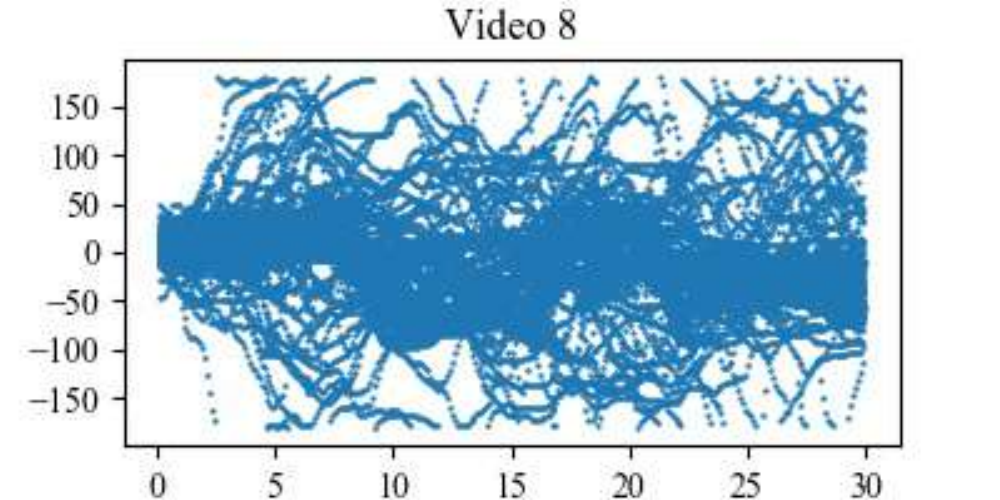}}\label{fig_third_case}
	
	\subfigure{\includegraphics[width=0.83in]{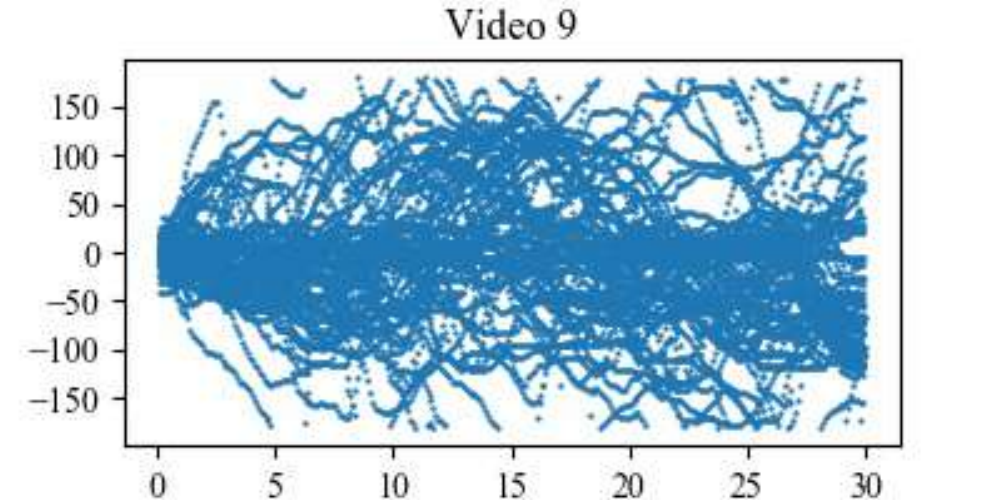}}\label{fig_first_case}
	\subfigure{\includegraphics[width=0.83in]{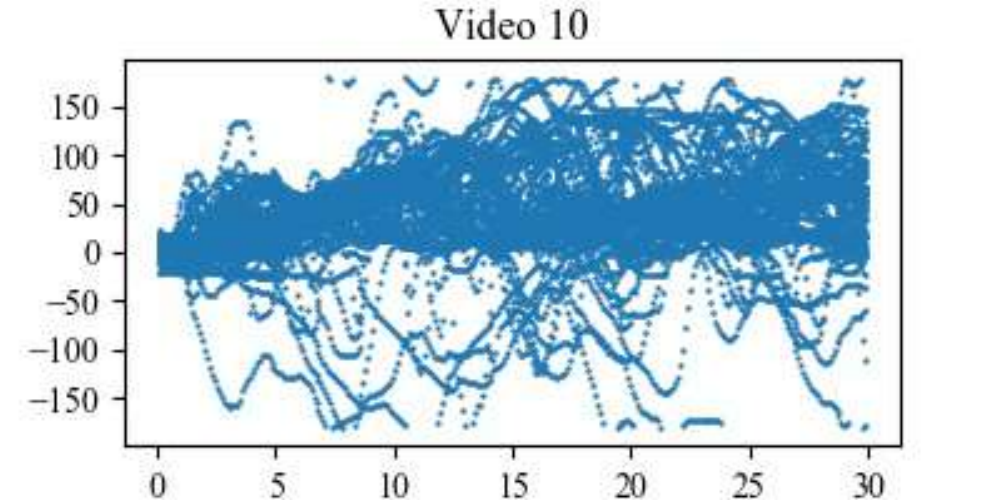}}\label{fig_second_case}
	\subfigure{\includegraphics[width=0.83in]{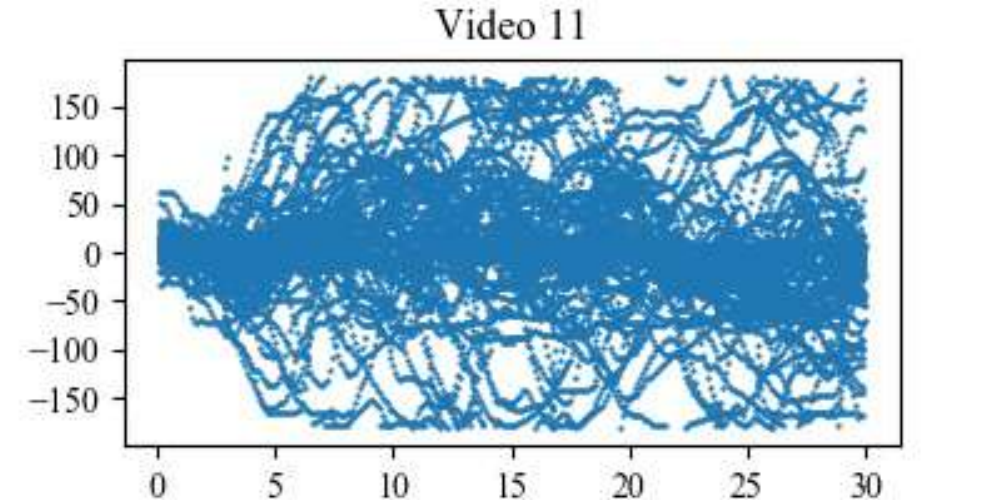}}\label{fig_third_case}
	\subfigure{\includegraphics[width=0.83in]{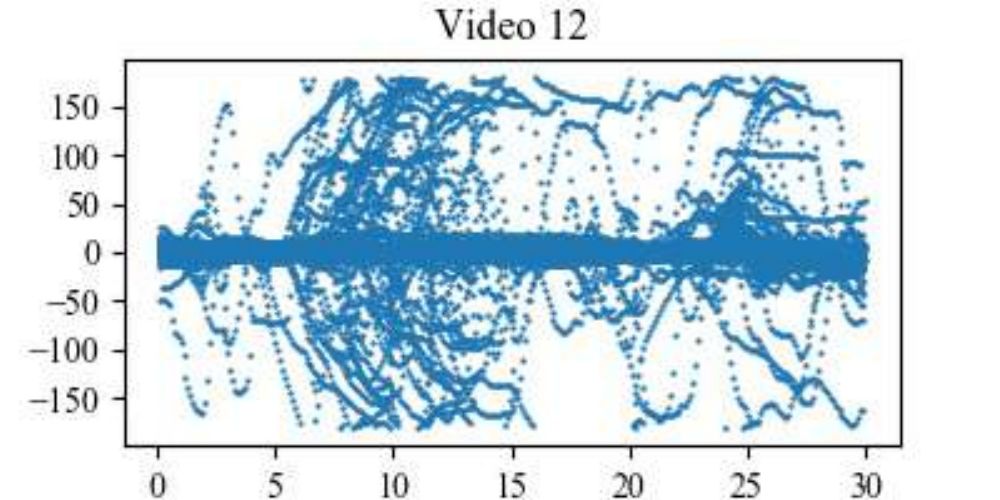}}\label{fig_third_case}
	
	\subfigure{\includegraphics[width=0.83in]{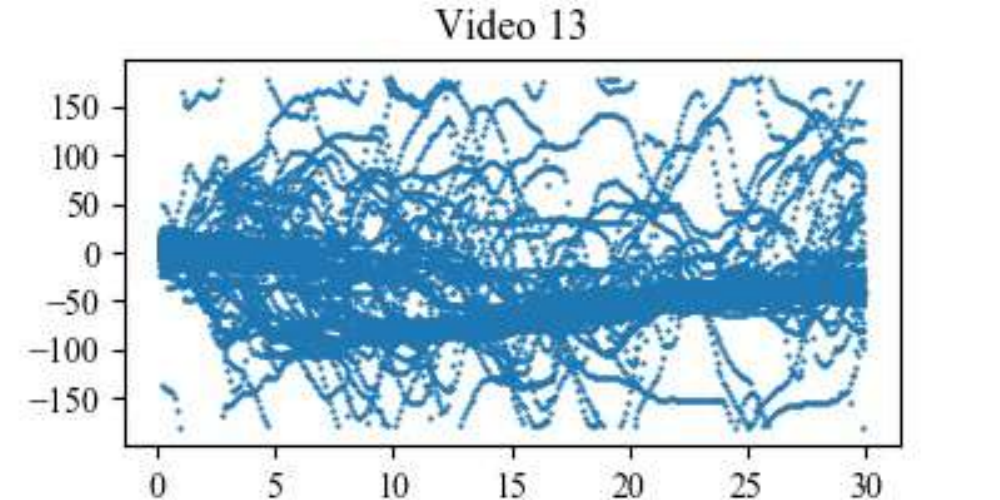}}\label{fig_first_case}
	\subfigure{\includegraphics[width=0.83in]{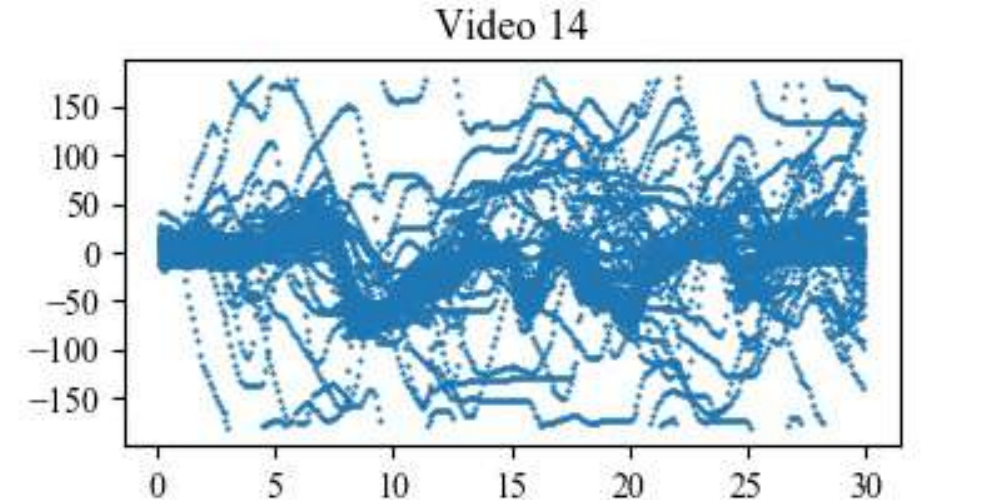}}\label{fig_second_case}
	\subfigure{\includegraphics[width=0.83in]{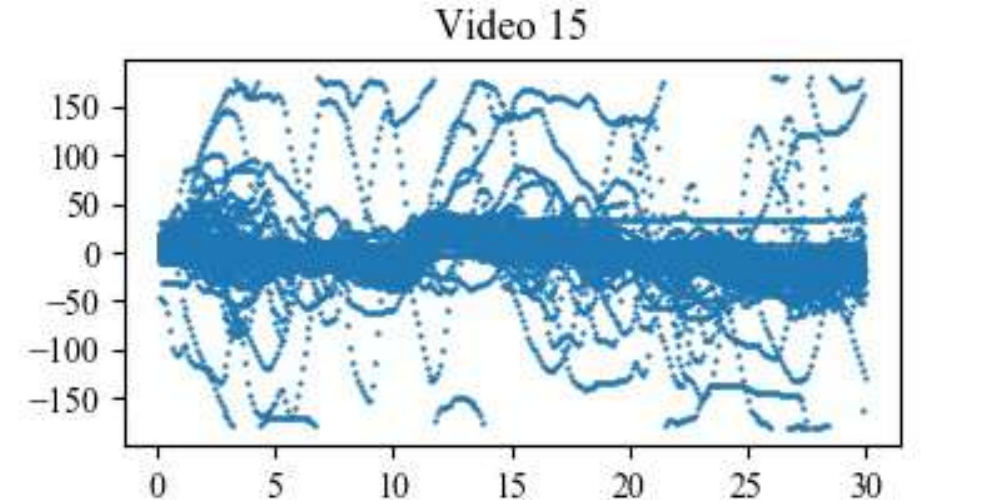}}\label{fig_third_case}
	\subfigure{\includegraphics[width=0.83in]{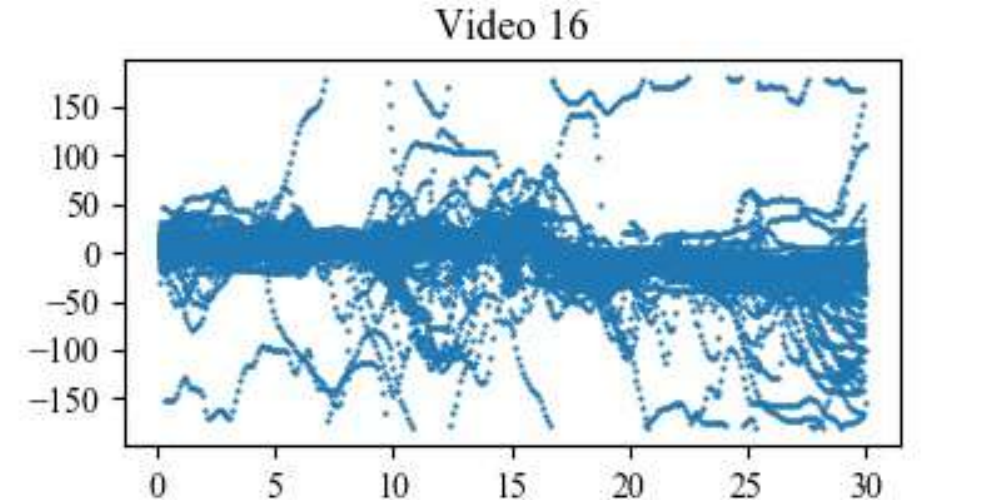}}\label{fig_third_case}
	\caption{$Y$ angle distribution of all VR users.}
	\label{basic_modules}
\end{figure}

\begin{figure}[!t]
	\centering
	\subfigure{\includegraphics[width=0.83in]{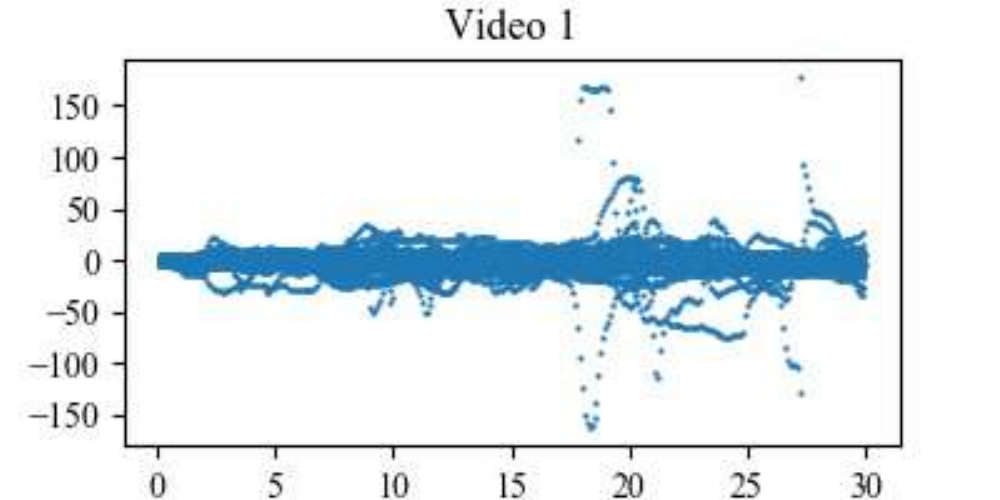}}\label{fig_first_case}
	\subfigure{\includegraphics[width=0.83in]{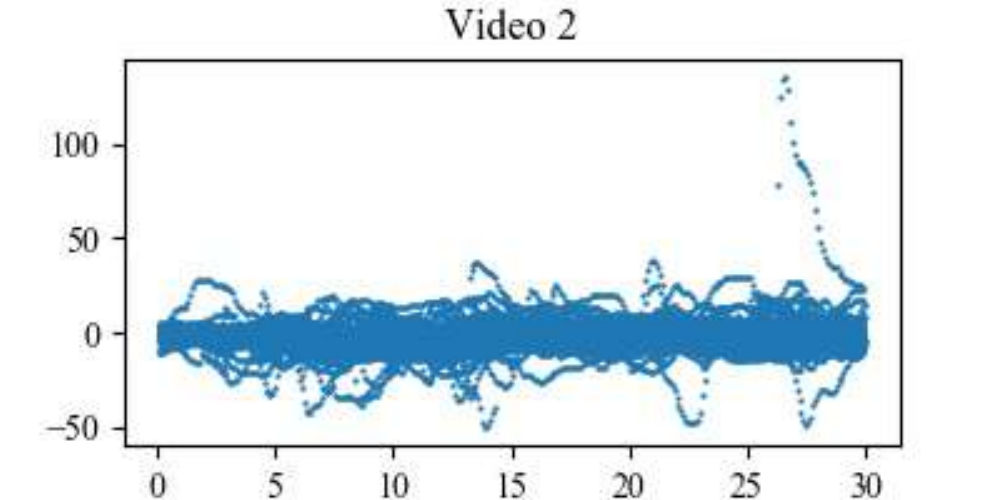}}\label{fig_second_case}
	\subfigure{\includegraphics[width=0.83in]{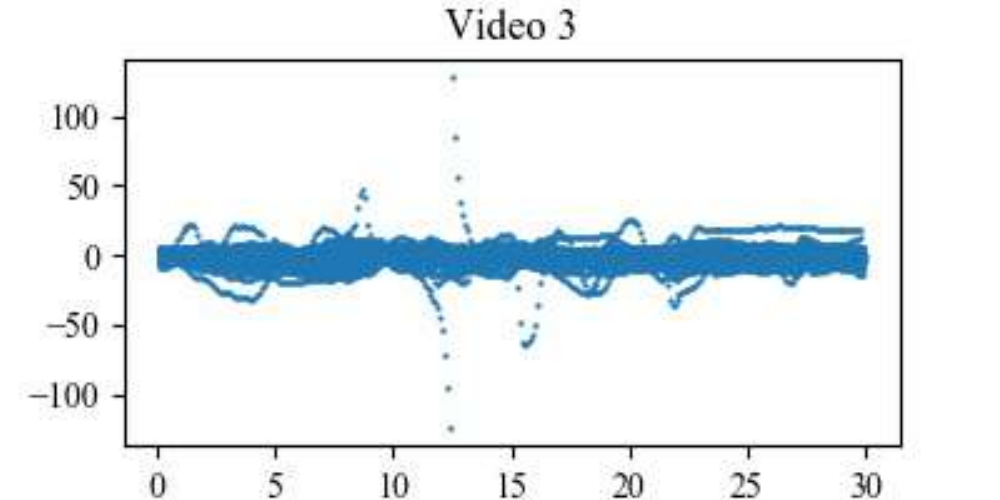}}\label{fig_third_case}
	\subfigure{\includegraphics[width=0.83in]{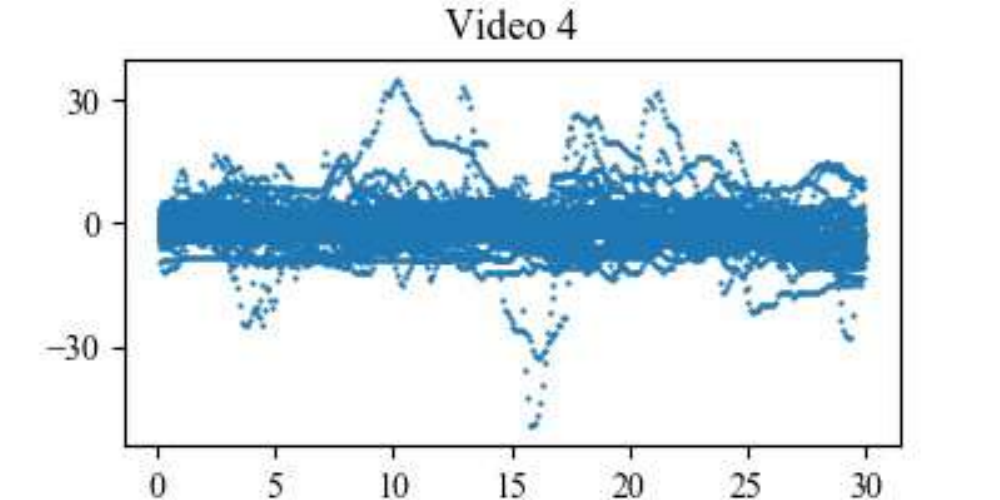}}\label{fig_third_case}
	
	\subfigure{\includegraphics[width=0.83in]{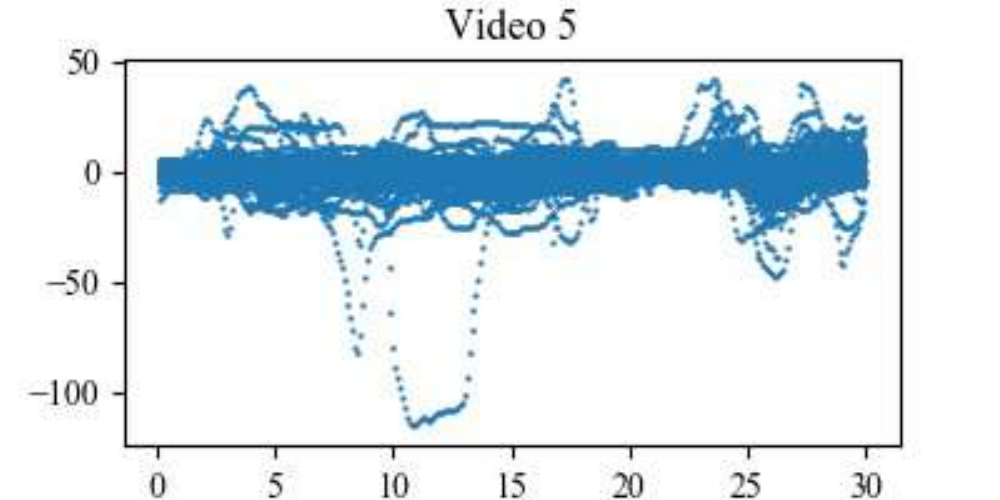}}\label{fig_first_case}
	\subfigure{\includegraphics[width=0.83in]{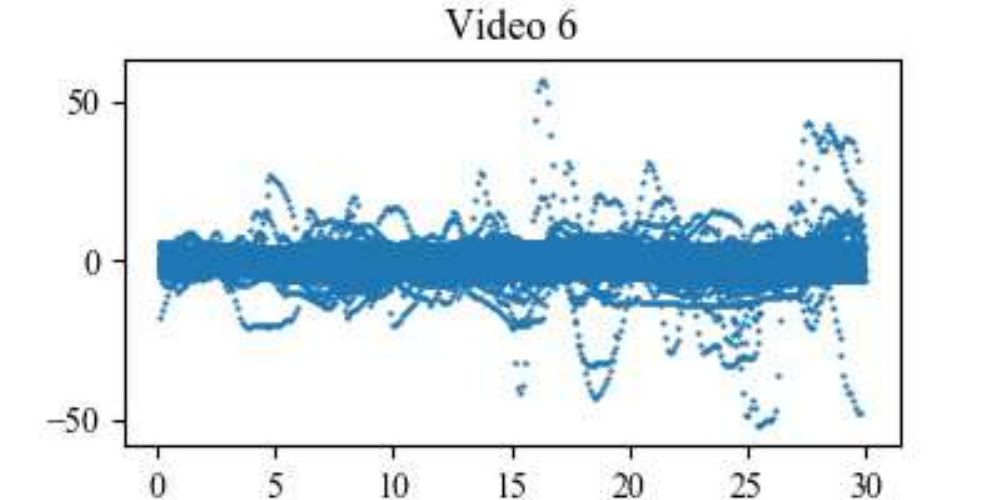}}\label{fig_second_case}
	\subfigure{\includegraphics[width=0.83in]{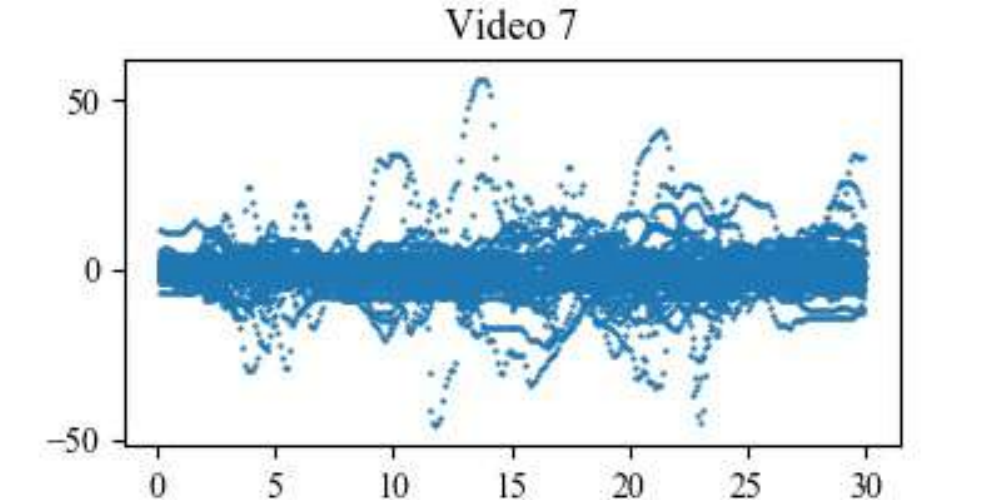}}\label{fig_third_case}
	\subfigure{\includegraphics[width=0.83in]{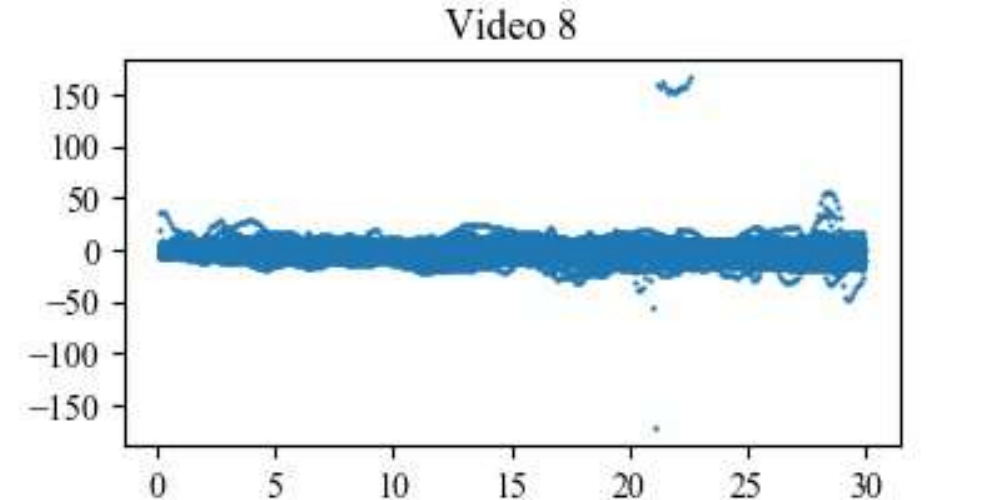}}\label{fig_third_case}
	
	\subfigure{\includegraphics[width=0.83in]{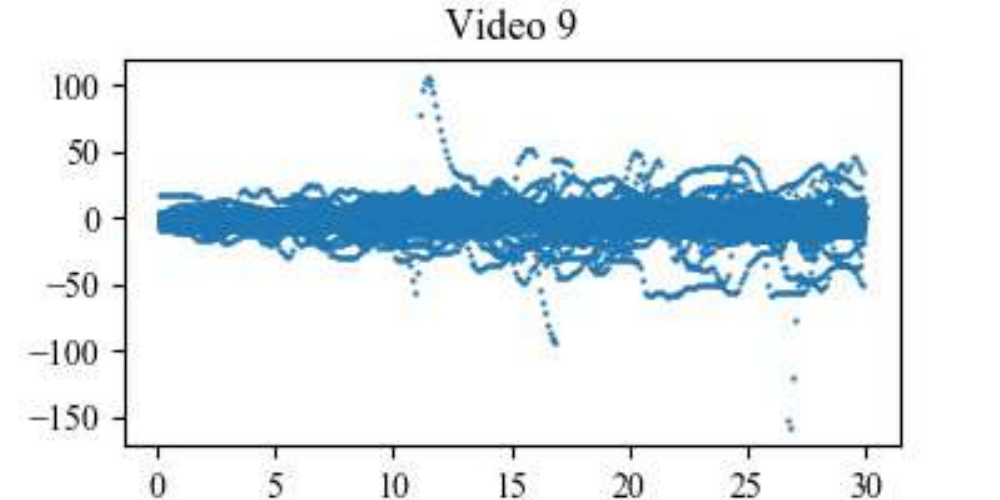}}\label{fig_first_case}
	\subfigure{\includegraphics[width=0.83in]{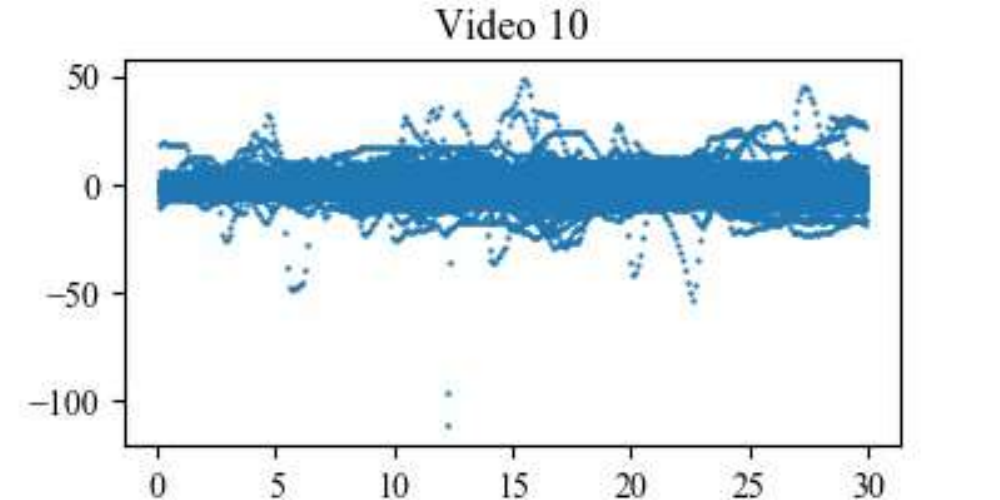}}\label{fig_second_case}
	\subfigure{\includegraphics[width=0.83in]{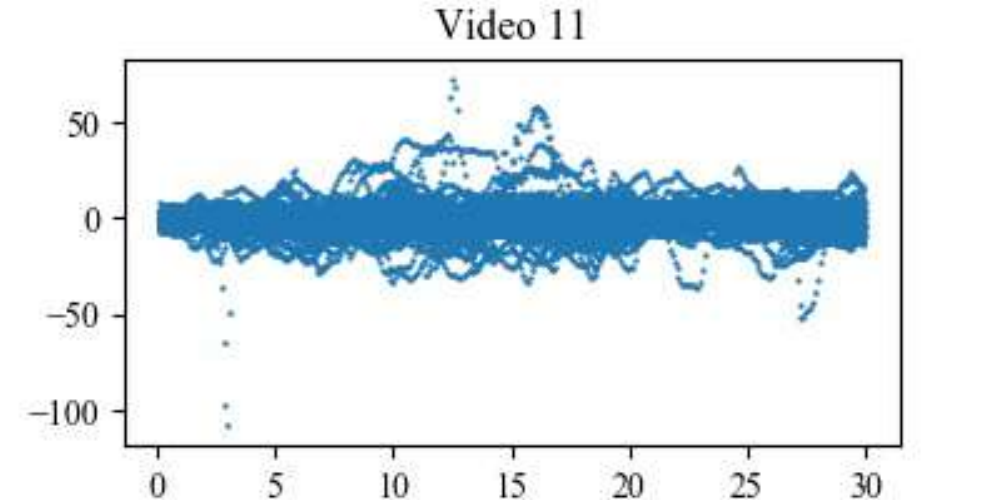}}\label{fig_third_case}
	\subfigure{\includegraphics[width=0.83in]{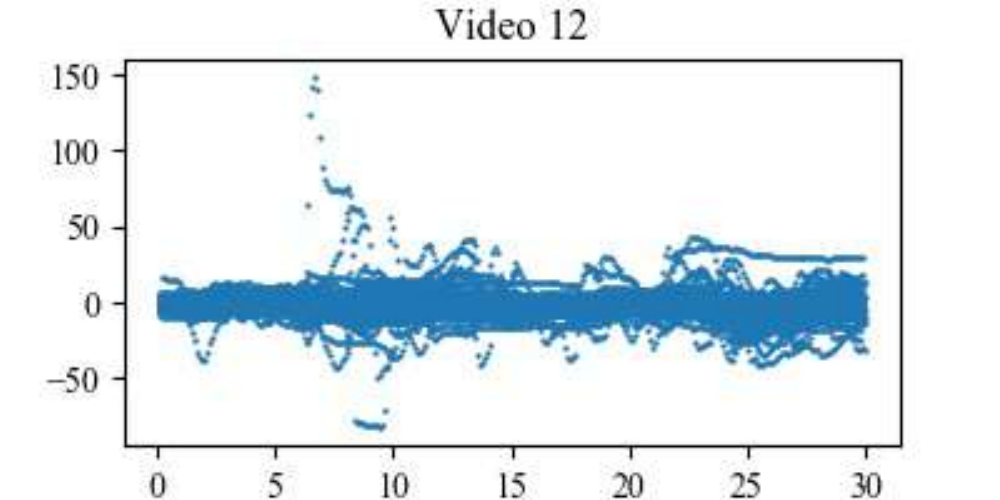}}\label{fig_third_case}
	
	\subfigure{\includegraphics[width=0.83in]{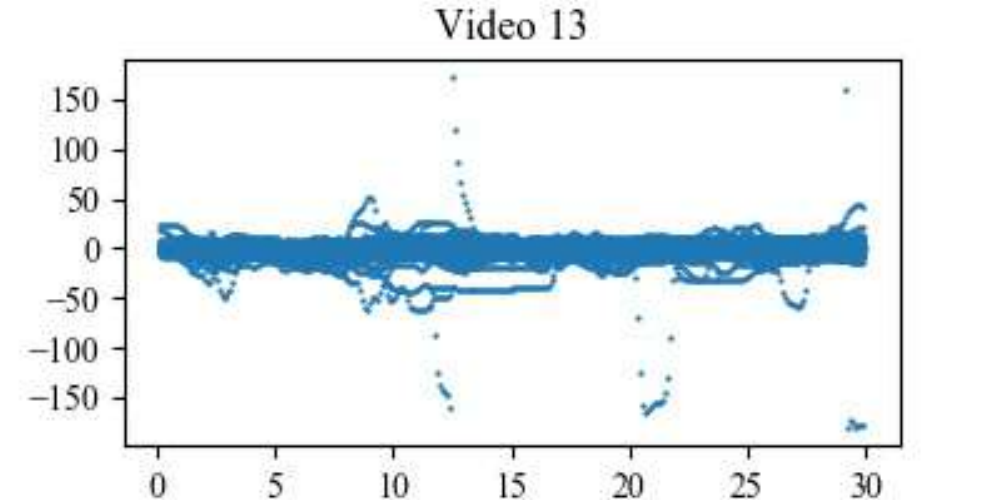}}\label{fig_first_case}
	\subfigure{\includegraphics[width=0.83in]{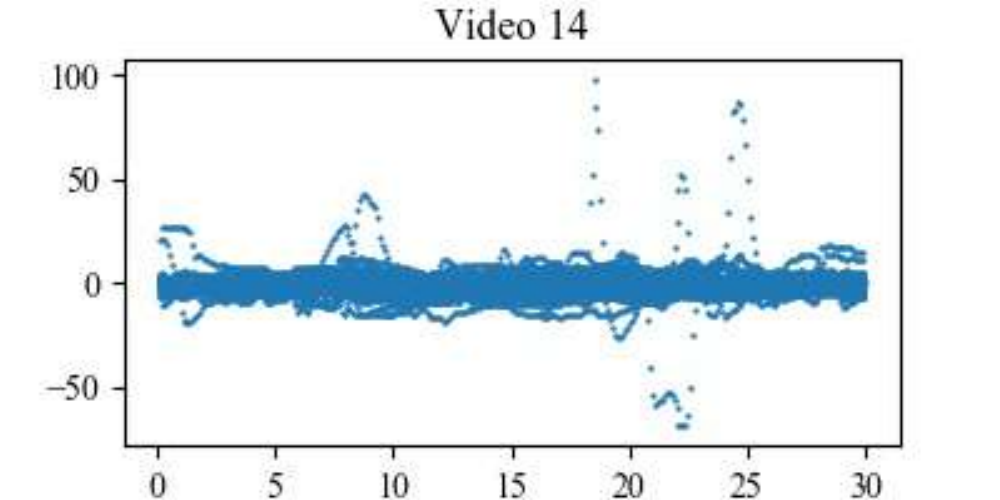}}\label{fig_second_case}
	\subfigure{\includegraphics[width=0.83in]{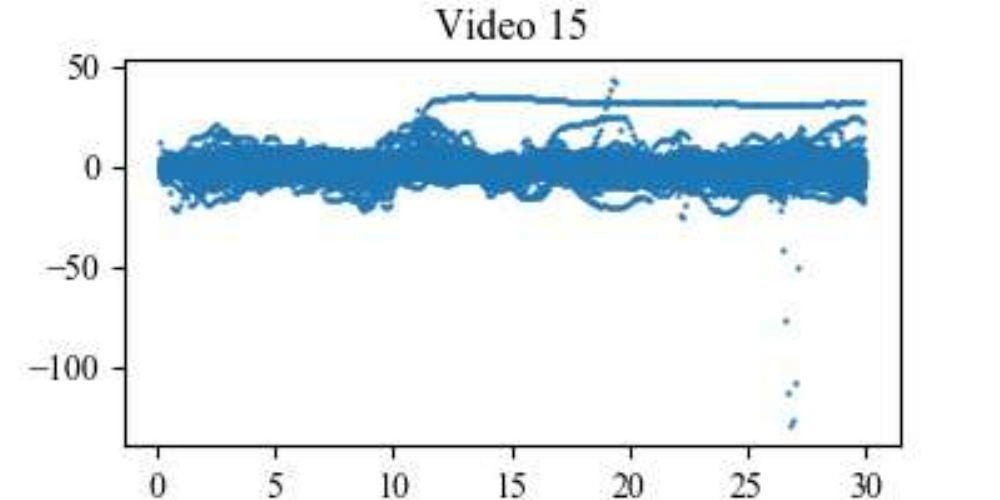}}\label{fig_third_case}
	\subfigure{\includegraphics[width=0.83in]{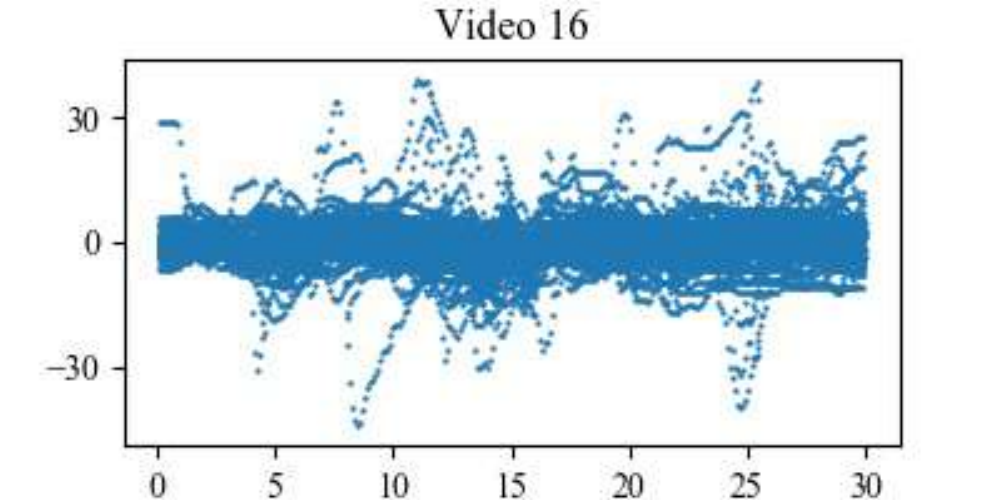}}\label{fig_third_case}
	\caption{$Z$ angle distribution of all VR users.}
	\label{basic_modules}
\end{figure}

\begin{figure}[!t]
    \centering
    \includegraphics[width=3.5 in]{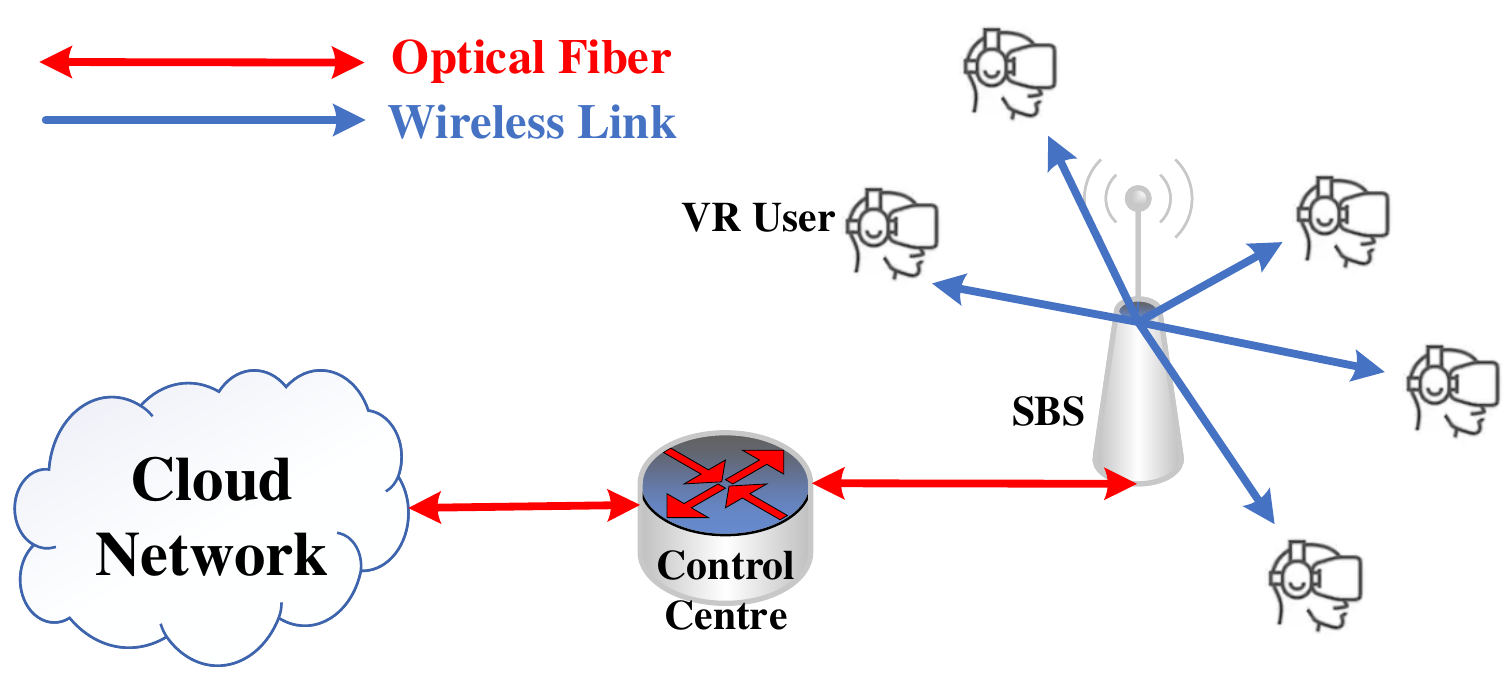}
    \caption{Wireless VR network.}
    \label{basic_modules}
\end{figure}

\section{System Model and Problem Formulation}
We consider a wireless VR system, where a small-cell base station (SBS) is connected to the core network through a fiber link to serve $K^{\rm{VR}}$ VR users via wireless links as shown in Fig. 8. The SBS is equipped with $M$ antennas and each VR user is equipped with a single antenna, respectively.

\subsection{Uplink Transmission Model}
At the $(t-1)$th time slot, the SBS will predict the viewpoint $\hat{V}_t^{k} = (\hat{X}_t^{k}, \hat{Y}_t^{k}, \hat{Z}_t^{k})$ of the $k$th VR user for the $t$th time slot. Then, at the $t$th time slot, the VR user will transmit its actual viewpoint ${V}_t^{k} = ({X}_t^{k}, {Y}_t^{k}, {Z}_t^{k})$ to the SBS via uplink transmission. The uplink transmission signal from the $k$th VR user to the SBS at the $t$th time slot can be denoted as
\begin{equation}
    \textbf{y}_{k,t}^{\rm{up}} = \textbf{u}_{k,t}^{H}{\textbf{h}}_{k,t}\textbf{x}_{k,t}^{\rm{up}} + \sum_{i =1, i\neq k}^{K^{\rm{VR}}}\textbf{u}_{k,t}^{H}{\textbf{h}}_{i,t}\textbf{x}_{i,t}^{\rm{up}} + \textbf{n}_{k,t}^{\rm{up}},
\end{equation}
where ${\textbf{h}}_{k,t}\in\mathbb{C}^{1\times M}$ is the uncorrelated Rayleigh fading channel vector between the $k$th VR user and the SBS at the $t$th time slot, $M$ is the number of antennas equipped at the SBS, and ${\alpha}$ is the large-scale fading coefficient. $\textbf{u}_{k,t}\in{\mathbb{C}^{1\times M}}$ is the beamforming vector at the SBS, which can be denoted as $\textbf{u}_{k,t} = \frac{{\textbf{h}}_{k,t}}{\|{\textbf{h}}_{k,t}\|}$ \cite{MRC1}. $\textbf{x}_{k,t}^{\rm{up}}\in\mathbb{C}^{M\times 1}$ is the transmit message of the $k$th VR user at the $t$th time slot. $\sum_{i =1, i\neq k}^{K}\textbf{u}_{k,t}^{H}{\textbf{h}}_{i,t}\textbf{x}_{i,t}^{\rm{up}}$ is the interference from other VR users at the $t$th time slot. In addition, $\textbf{n}_{k,t}^{\rm{up}}\sim \mathcal{CN}(0, \sigma_{k}^{2}\textbf{I}_{M})$ is the additive white Gaussian noise at the SBS at the $t$th time slot. 

Furthermore, at the $t$th time slot, the data rate between the $k$th VR user and the SBS can be written as
\begin{equation}
    R_{k,t}^{\rm{up}} = \log_{2}\left|\textbf{I} + \frac{|\textbf{u}_{k,t}^{H}{\textbf{h}}_{k,t}|^2}{\sum\limits_{i =1, i\neq k}^{K^{\rm{VR}}}|\textbf{u}_{k,t}^{H}{\textbf{h}}_{i,t}|^2+ \sigma_{k}^{2}\textbf{I}_{M}}\right|.
\end{equation}
To guarantee the successful uplink transmission, the uplink transmission rate should larger than a threshold $R_{\rm{th}}^{\rm{up}}$, namely, $R_{k,t}^{\rm{up}}\geq R_{\rm{th}}^{\rm{up}}$. However, it is possible that the uplink transmission rate of the $k$th VR user is smaller than the threshold because of the interference or the poor channel state information. 

To guarantee the reliability of uplink transmission, we consider the proactive retransmission scheme. If the actual viewpoint from the VR user is successfully transmitted via the uplink transmisson, the SBS will send an ACK feedback, otherwise, it will send a NACK feedback.

According to the proactive scheme, the $k$th VR user will repeat the uplink transmission in consecution transmission time intervals (TTIs) with a maximum number of $K_{\rm{re}}$ repetitions, but can receive the feedback after each repetition. The $k$th VR user is allowed to stop repetitions once receiving the positive feedback (ACK). We assume that the processing time of the received viewpoint and feedback time at the SBS are one transmission time interval (TTI), respectively. For example, when $K_{\rm{re}}=8$, as shown in Fig. 9, we can observe that the $k$th VR user is able to receive the 1st feedback in $4\rm{TTIs}$ after the 1st repetition, which means that the minimum round trip time is $4\rm{TTIs}$. Nevertheless, if the $k$th VR user cannot obtain the ACK at the first round trip time, it will continue waiting for the ACK until $(K_{\rm{re}}+3)\rm{TTIs}$. However, if the $k$th VR user cannot obtain the ACK during the initial transmission, it needs to continue repetitions until either it receives ACK from the SBS, or the latency is larger than the uplink transmission latency threshold. If the 1st successful uplink transmission of the $k$th VR user occurs in the $l$th repetition during the first round trip, the uplink latency of the $k$th VR user can be computed as 
\begin{equation}
    T_{k,l} = (l + 3)\rm{TTIs},
\end{equation}
Furthermore, the latency after $m$ round trips for the Proactive scheme with a maximum $K_{\rm{re}}$ repetitions can be derived as
\begin{align}
    T_{k,l}^{m} &= (m-1)T_{k,K_{\rm{re}}} + T_{k,l} \\ \nonumber
    &= [(m-1)(K_{\rm{re}} + 3) + (l + 3)]\rm{TTIs},
\end{align}
where $(m-1)T_{k,K_{\rm{re}}}$ means that the uplink transmissions in former $(m-1)$ round trips are not successful, and $T_{k,l}$ denotes the successful uplink retransmission in the final $m$th round trip given in (3).

\begin{figure}[!h]
    \centering
    \includegraphics[width=3.5 in]{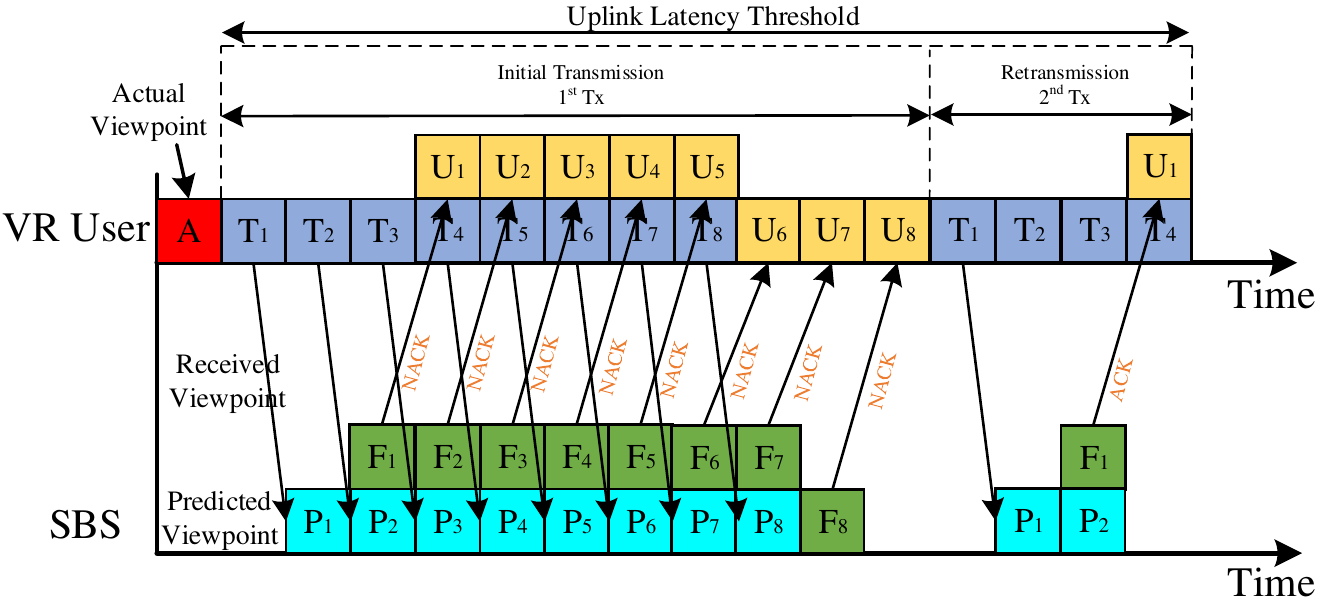}
    \caption{Proactive retransmission scheme.}
    \label{basic_modules}
\end{figure}

\subsection{Viewpoint Prediction Methods}
When the VR user enjoys the VR video frame, the viewpoint has three degrees of freedom (pitch, yaw, and roll) and is determined by the rotation angles in $X$, $Y$ and $Z$ axis. Therefore, predicting the viewpoint of the VR user is equal to predicting the $X$, $Y$ and $Z$ angles.
We consider a sliding window to predict the viewpoint of the VR user over time, which is shown in Fig. 10. According to Fig. 10, the future viewpoint of the VR user is predicted based on the current and past rotation status. We assume that the pitch, yaw and roll angles of the VR user at the $t$th time slot are $X_t$, $Y_t$, and $Z_t$, respectively. Furthermore, we use $\textbf{X}_{t:(t+d)} = (X_{t},X_{t+1},...,X_{t+d})$, $\textbf{Y}_{t:(t+d)} = (Y_{t}, Y_{t+1},...,Y_{t+d})$ and $\textbf{Z}_{t:(t+d)} = (Z_{t}, Z_{t+1},...,Z_{t+d})$ to denote the continuous viewpoints in $X$, $Y$ and $Z$ angles from the $t$th time slot to the $(t+d)$th time slot.

\begin{figure}[!h]
    \centering
    \includegraphics[width=3.5 in]{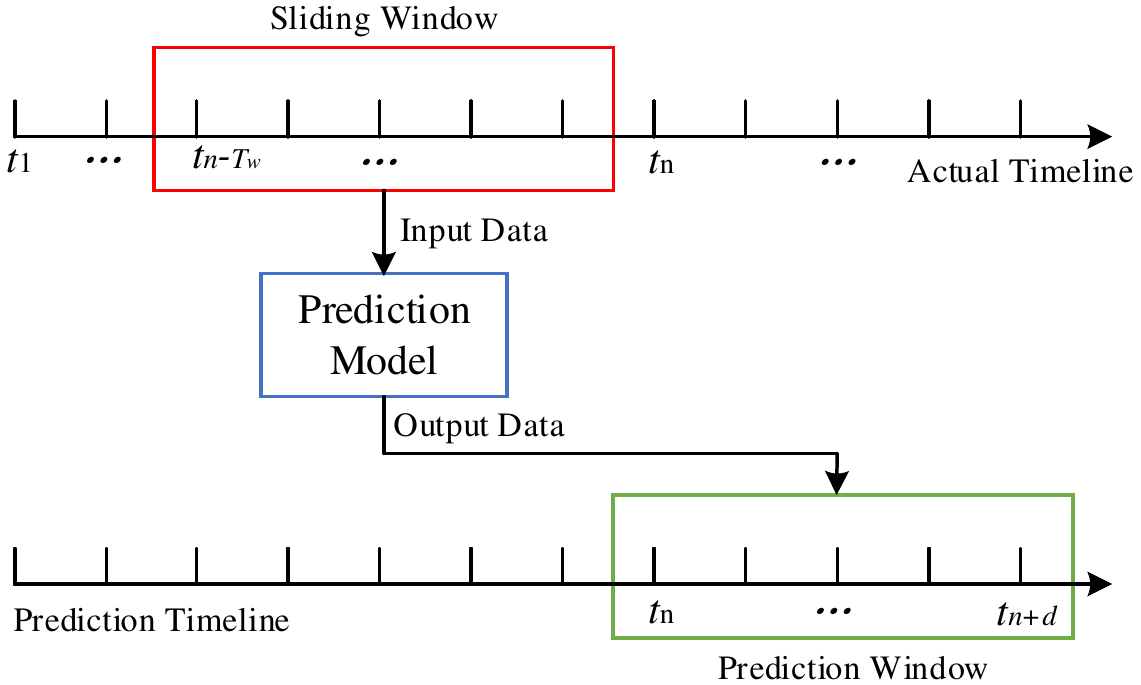}
    \caption{Sliding window.}
    \label{basic_modules}
\end{figure}

To predict the future viewpoint $(X_{t+d}, Y_{t+d}, Z_{t+d})$ at the $t$th time slot, we use previous viewpoint $\textbf{X}_{(t-T_w):t}$, $\textbf{Y}_{(t-T_w):t}$ and $\textbf{Z}_{(t-T_w):t}$, where $T_w$ is the size of the sliding window. Then, the predicted viewpoint at the $(t+d)$th time slot can be presented as
\begin{equation}
    \hat{X}_{t+d} = f_{x,t+d}(\textbf{X}_{(t-T_w):t}),
\end{equation}
\begin{equation}
    \hat{Y}_{t+d} = f_{y,t+d}(\textbf{Y}_{(t-T_w):t}),
\end{equation}
\begin{equation}
    \rm{and}~\hat{Z}_{t+d} = f_{z,t+d}(\textbf{Z}_{(t-T_w):t}),
\end{equation}
where $f_{x,t+d}(.)$, $f_{y,t+d}(.)$, and $f_{z,t+d}(.)$ are prediction function.

To predict the viewpoint of the VR user accurately, we consider two learning algorithms, namely, offline learning and online learning.

\subsubsection{Offline Learning}
In the offline learning algorithms, we propose three methods, which are trained $n$-order Linear Regression (LR), Neural Network (NN), and Recurrent Neural Network (RNN) based on Long-short Term Memory(LSTM)/Gated Recurrent Unit (GRU) architecture to predict the viewpoint of VR users. Through dividing the VR dataset into training and testing dataset, the VR user data in the training dataset are used to train the models for three offline methods, where the trained models are used to predict the viewpoint of the VR user directly.

\subsubsection{Online Learning}
In the online learning algorithms, we still use $n$-order LR, NN and LSTM/GRU  algorithms. Meanwhile, we use Mean Square Error (MSE) \cite{MSE} as cost function in the training step to update the parameters in the online learning model, and predict the viewpoint of new VR users. The MSE of the VR users at the $t$th time slot can be presented as
\begin{equation}
    \text{MSE}_{t} = \frac{1}{K^{\rm{VR}}}\sum\limits_{k=1}^{K^{\rm{VR}}}(\hat{V}_{t}^{k}-V_{t}^{k})^2.
\end{equation}
At the $t$th time slot, the SBS will predict the viewpoint of the VR user for the $(t+1)$th time slot. At the $(t+1)$th time slot, the VR user will transmit the actual viewpoint to the SBS via uplink transmission. Through comparing it with the predicted viewpoint, the SBS will further update the trained learning model to improve the prediction accuracy.

\subsection{Rendering and Downlink Transmission Model}
When the future viewpoint of the VR user is predicted via offline or online learning algorithms, the SBS will render the predicted viewpoint and transmit it to the VR user through downlink transmission in advance. Therefore, the VR interaction latency can be reduced \cite{xiaonan}. In this paper, we mainly focus on prediction and uplink retransmission in the wireless VR system, which can be easily integrated into the rendering and downlink transmission in our previous work \cite{xiaonan}.

\subsection{Problem Formulation}
For the viewpoint prediction, we use offline and online learning algorithms to minimize the average prediction error of VR users, the optimization problem can be formulated as
\begin{equation}
    \min \frac{1}{T_{i}^{\rm{tot}}\widetilde{N}_i}\sum_{t=1}^{T_{i}^{\rm{tot}}}\sum_{k=1}^{\widetilde{N}_i}(\hat{V}_{t}^{k}-V_{t}^{k})^2,
\end{equation}
where $\widetilde{N}_i$ is the number of the VR users watching the $i$th VR video, and $T_{i}^{\rm{tot}}$ is the total time slots of the $i$th VR video.

\section{Learning Algorithms for Viewpoint Prediction}
In the offline learning, we directly use the trained $n$-order LR, NN and LSTM/GRU network to predict the viewpoint of the VR user in continuous time slots. However, for the online learning, the VR user will deliver the actual viewpoint to the SBS via uplink transmission in real-time to further update the models in the NN and LSTM/GRU learning algorithms and the input of the sliding window, which can improve the prediction accuracy. If the actual viewpoint at a specific time slot has not been successfully delivered to the SBS, the learning algorithms will predict the viewpoint in the next time slot based on the models trained in the previous time slots, and the input of the sliding window at the current time slot is set to be null. 

\subsection{Offline Learning Algorithm}
According to Fig. 3, VR dataset contains dozens of VR users enjoy each VR video, and there are 16 VR videos and 969 VR user samples. To train the learning model, we split the data samples of the VR dataset into the training and the testing datasets. The training dataset is used to train the learning model, and the testing dataset is used to validate it on data it has never seen before. The classic approach is to do a simple 80$\%$-20$\%$ \cite{split}, which means that we randomly select 80$\%$ data samples of the dataset to construct training dataset, while the remaining 20$\%$ data samples of the dataset are the testing dataset. However, with a simple 80-20 split, there is a possibility of high bias if we have limited data. More importantly, we will miss some important information about the data samples which are not used for training, which is able to get good or bad performance only due to chance. To ensure that each data sample from the original dataset has the chance of appearing in the training and testing dataset, we use K Cross Validation \cite{Kcross}.

Through using the K Cross Validation to train the learning models in the VR dataset, each VR user sample has the opportunity of being tested. We split the VR dataset into $K_{\rm{cross}}$ datasets: one dataset is used for validation, and the remaining $(K_{\rm{cross}}-1)$ datasets are merged into a training dataset for prediction learning model evaluation \cite{Cross}. In our VR dataset, there are 16 different VR videos. According to Fig 5, 6 and 7, the viewpoint distribution of VR users in each VR video are different. Therefore, for each VR video, we can use its corresponding VR user samples to train a viewpoint prediction learning model. However, if the number of the VR videos increases, training one model for one VR video may cost much more energy and occupy much more computation resource and memory of the SBS. Therefore, in order to evaluate the generality of the trained models, we propose two viewpoint prediction learning models, namely, one for single VR video, and the other for all VR videos. The detailed K Cross Validation for the proposed two viewpoint prediction schemes are introduced as follows:

(a) \textbf{One Model for One VR Video:} For each VR video, there are dozens of VR user samples, and we assume that the number of the VR user samples of the $k$th VR video is $\widetilde{N}_{k}$. We split these dozens of VR user samples into $K_{\rm{cross}}$ datasets, where the number of VR user samples in each sub dataset is $\widetilde{N}_{k}/K_{\rm{cross}}$, and randomly select ($K_{\rm{cross}}-1$) sub datasets to train the learning model and one sub dataset to test the trained learning model. Through $K_{\rm{cross}}$ times training and testing, we can obtain the average prediction error of the $K$-folder cross validation.

(b) \textbf{One Model for All VR Videos:} For all 16 VR videos and 969 VR user samples, we split these VR user samples according to the index of VR video. Therefore, there are 16 VR sub datasets. At each training round of the $K$ cross validation, $16/K_{\rm{cross}}$ VR sub datasets are used for testing, and the remaining $(16-16/K_{\rm{cross}})$ VR sub datasets are merged into a training VR sub dataset to train the viewpoint prediction model. For example, in Fig. 11, we consider 4 cross validation, namely, $K_{\rm{cross}}=4$. At each training iteration, 12 VR sub datasets will be randomly selected to train the learning model, and the remaining 4 VR sub dataset will be used to test the trained learning model. After 4 training iterations, we can calculate the average prediction error of these 4 trained learning models.

\begin{figure}[!h]
    \centering
    \includegraphics[width=3.5 in]{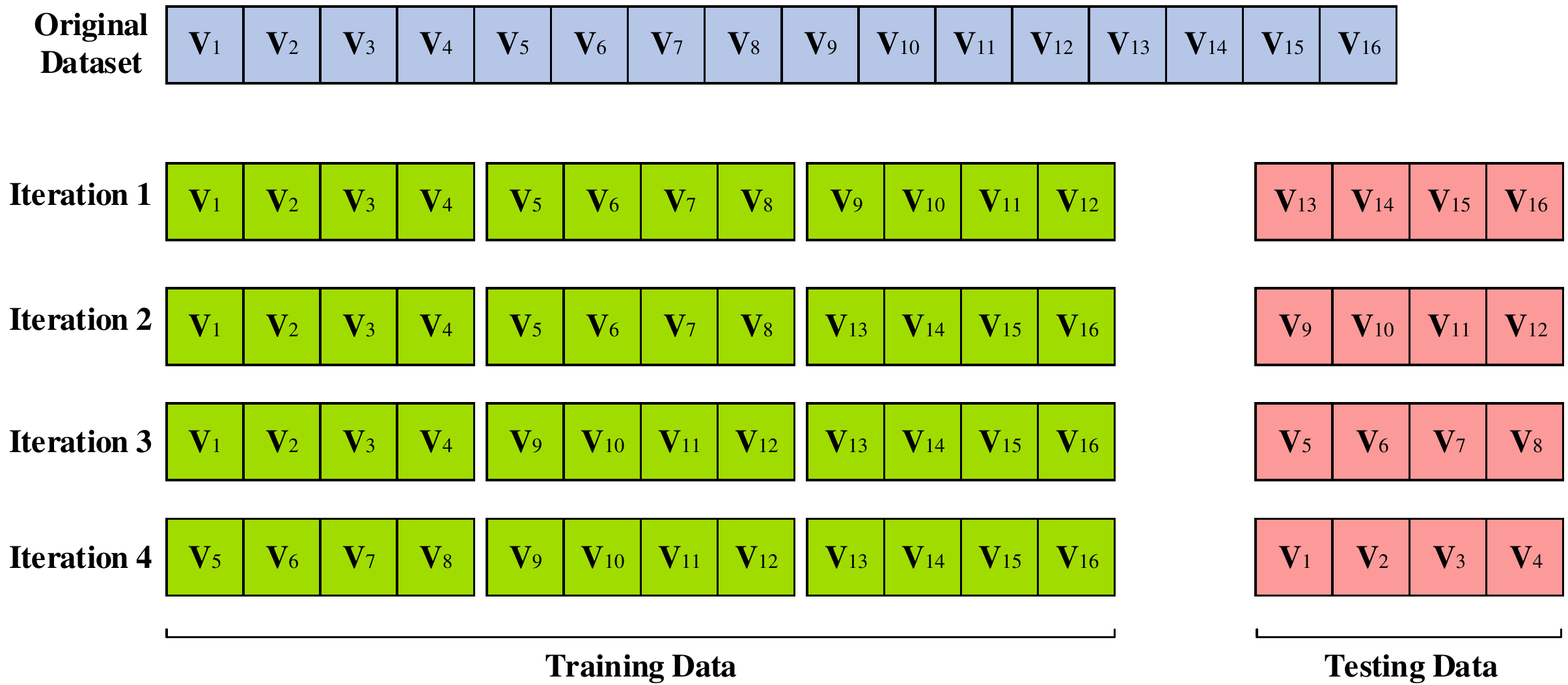}
    \caption{4 cross validation for training learning models.}
    \label{basic_modules}
\end{figure}

\subsection{Online Learning Algorithm}
In the online learning algorithms, the learning model will first be trained via the training dataset through the K Cross Validation described in Section IV-A. Then, for the VR user samples in the testing dataset, at each time slot, each VR user will update its actual viewpoint to the SBS through uplink transmission. The online learning algorithms are introduced in detail as follows.

\subsubsection{$n$-order Linear Regression}
$n$-order LR algorithm uses the least square function to model the nolinear relationship between the input sliding window and the predicted viewpoint. It is able to fit the nonlinear relationship between the input and output, and can be written as
\begin{equation}
    \hat{V} = \textbf{W}^{\rm{LR}}\textbf{g}^{H} + {b}^{\rm{LR}},
\end{equation}
where $\textbf{W}^{\rm{LR}} = [w_1^{\rm{LR}},w_2^{\rm{LR}},...,w_n^{\rm{LR}}]$ and ${b}^{\rm{LR}}$ are parameters of the $n$-order LR model. In (10), $\textbf{g}=[\hat{\textbf{g}},\hat{\textbf{g}}^2,...,\hat{\textbf{g}}^n]$ is the input of the $n$-order LR, where $\hat{\textbf{g}} = (\textbf{X}_{(t-T_w):t},\textbf{Y}_{(t-T_w):t},\textbf{Z}_{(t-T_w):t})$ is the vector which includes the $X$, $Y$ and $Z$ viewing angles in $T_w$ time slots. $\hat{V}_{t+1}=(\hat{X}_{t+1}, \hat{Y}_{t+1}, \hat{Z}_{t+1})$ is the predicted viewing angles for the $(t+1)$th time slot. The loss function of the $n$-order LR can be calculated as
\begin{equation}
    \mathcal{L}_t^{\rm{LR}} = \frac{1}{K^{\text{VR}}}\sum\limits_{k=1}^{K^{\text{VR}}}(V_{t}^{k}-\hat{V}_{t}^{k})^2.
\end{equation}
Through gradient descent method \cite{SGD1}, the parameters $\bm{\theta}^{\rm{LR}}=\{\textbf{W}^{\rm{LR}}, {b}^{\rm{LR}}\}$ can be updated as
\begin{equation}
    \bm{\theta}_{t+1}^{\rm{LR}} = \bm{\theta}_{t}^{\rm{LR}}-\Delta\mathcal{L}_t^{\rm{LR}}(\bm{\theta}_{t}^{\rm{LR}}),
\end{equation}
where $\Delta \mathcal{L}^{\rm{LR}}(.)$ is the gradient of the loss function. The proposed Proactive retransmission scheme integrated into the online $n$-order LR is illustrated in Algorithm 1.

\begin{figure}[!h]
    \centering
    \includegraphics[width=3.5 in]{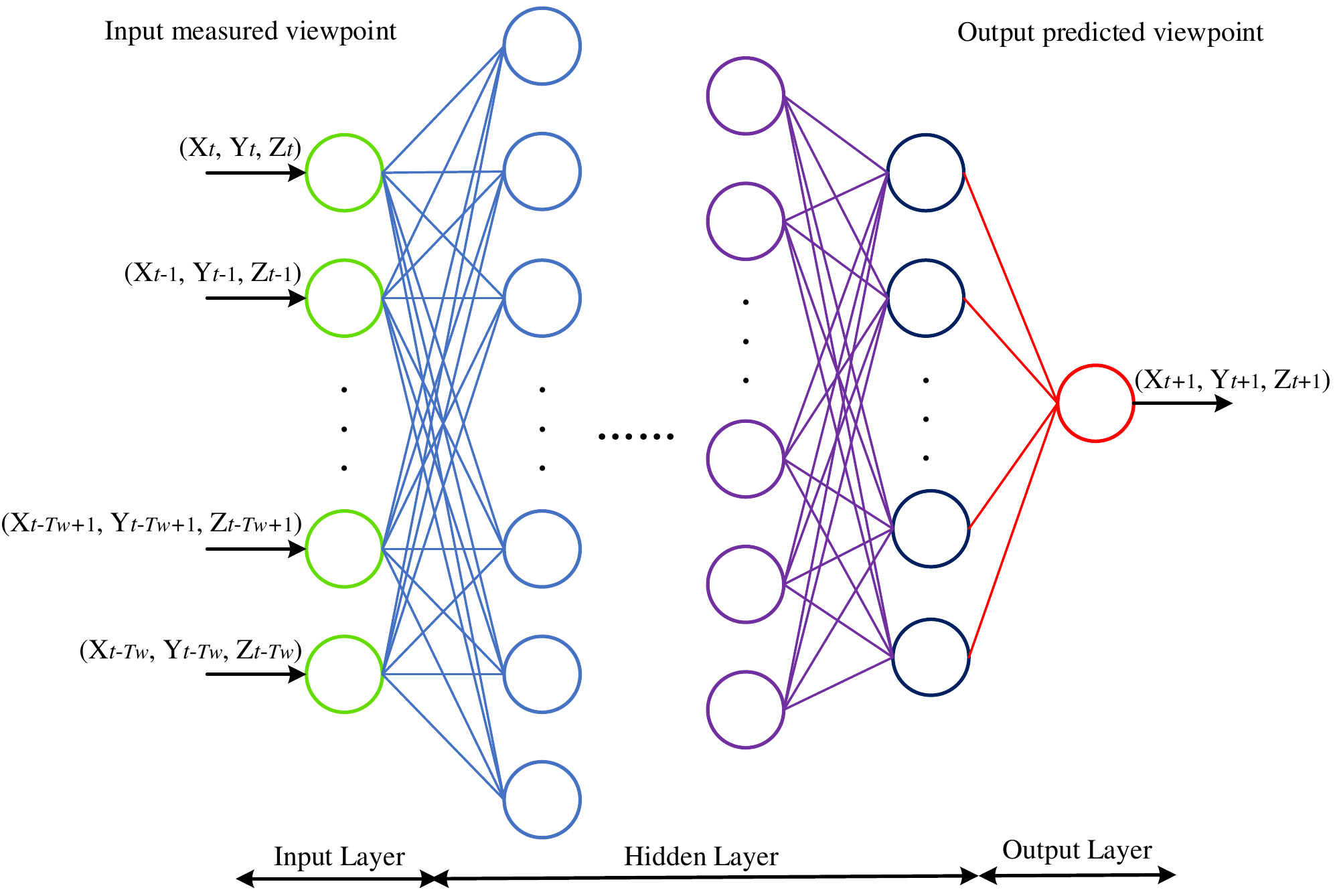}
    \caption{Proposed multi-layer NN architecture.}
    \label{basic_modules}
\end{figure}

\subsubsection{Neural Network}
In the $L$-layer NN shown in Fig. 12, we assume that $\Theta^{\rm{NN}}=\{\bm{\theta}_1^{\rm{NN}}, \bm{\theta}_2^{\rm{NN}},...,\bm{\theta}_L^{\rm{NN}}\}$ contains $L$ sets of parameters, and the parameters at the $l$th ($1\leq   l\leq L$) layer can be denoted as $\bm{\theta}_l^{\rm{NN}}=\{\bm{W}_l^{\rm{NN}}, \bm{b}_l^{\rm{NN}}\}$, where $\bm{W}_l^{\rm{NN}}$ and $\bm{b}_l^{\rm{NN}}$ are the neurons' weights and bias vector at the $l$th layer. A feedforward NN with $L$ layers describes a mapping $f^{\rm{NN}}(\textbf{r}^{\rm{NN}}, \bm{\theta}^{\rm{NN}})$, where $\textbf{r}^{\rm{NN}}$ is the input vector. We can obtain the output of the NN through $L$ iterative processing steps, and the output of the $l$th layer in NN can be written as
\begin{equation}
    \textbf{r}_l^{\rm{NN}} = f_l^{\rm{NN}}(\textbf{r}_{l-1}^{\rm{NN}};\bm{\theta}_l^{\rm{NN}}), l = 1, 2,...,L,
\end{equation}
where $f_l^{\rm{NN}}(\textbf{r}_{l-1}^{\rm{NN}};\bm{\theta}_l^{\rm{NN}})$ is the mapping function calculated by the $l$th NN layer.

At the $t$th time slot, we input the historical viewpoint of the VR user and obtain the predicted viewpoint via the feedforward function in the $L$-layer NN. Then, we use the MSE criterion among the predicted viewpoint and the actual viewpoint of the $(t+1)$th time slot to compute the loss of the NN, which can be denoted as
\begin{equation}
    \mathcal{L}_{t,l}^{\rm{NN}}(\bm{\theta}_{t,l}^{\rm{NN}}) = \|\bm{\phi}_{t,l}^{\rm{NN}}-\hat{\bm{\phi}}_{t,l}^{\rm{NN}}\|_2,
\end{equation}
where $\bm{\phi}_{t,l}^{\rm{NN}}$ is the desired output of the $l$th layer in NN, $\hat{\bm{\phi}}_{t,l}^{\rm{NN}}$ is the dependence of the NN's output to the $l$th layer's parameters. To minimize the loss function, we adopt the backpropagation method based on stochastic gradient descent (SGD) \cite{SGD2}. The parameters at the $l$th layer can be updated as
\begin{equation}
\bm{\theta}_{t+1,l}^{\rm{NN}} = \bm{\theta}_{t,l}^{\rm{NN}} - \lambda^{\rm{NN}}\Delta \mathcal{L}_{t.l}^{\rm{NN}}(\bm{\theta}_{t,l}^{\rm{NN}}),
\end{equation}
where $\lambda^{\rm{NN}}\in(0,1]$ denotes the learning rate of the NN and $\Delta \mathcal{L}^{\rm{NN}}(.)$ is the gradient of the loss function. The proposed Proactive retransmission scheme integrated into the online NN is presented in Algorithm 1.

\begin{figure}[!h]
    \centering
    \includegraphics[width=3.5 in]{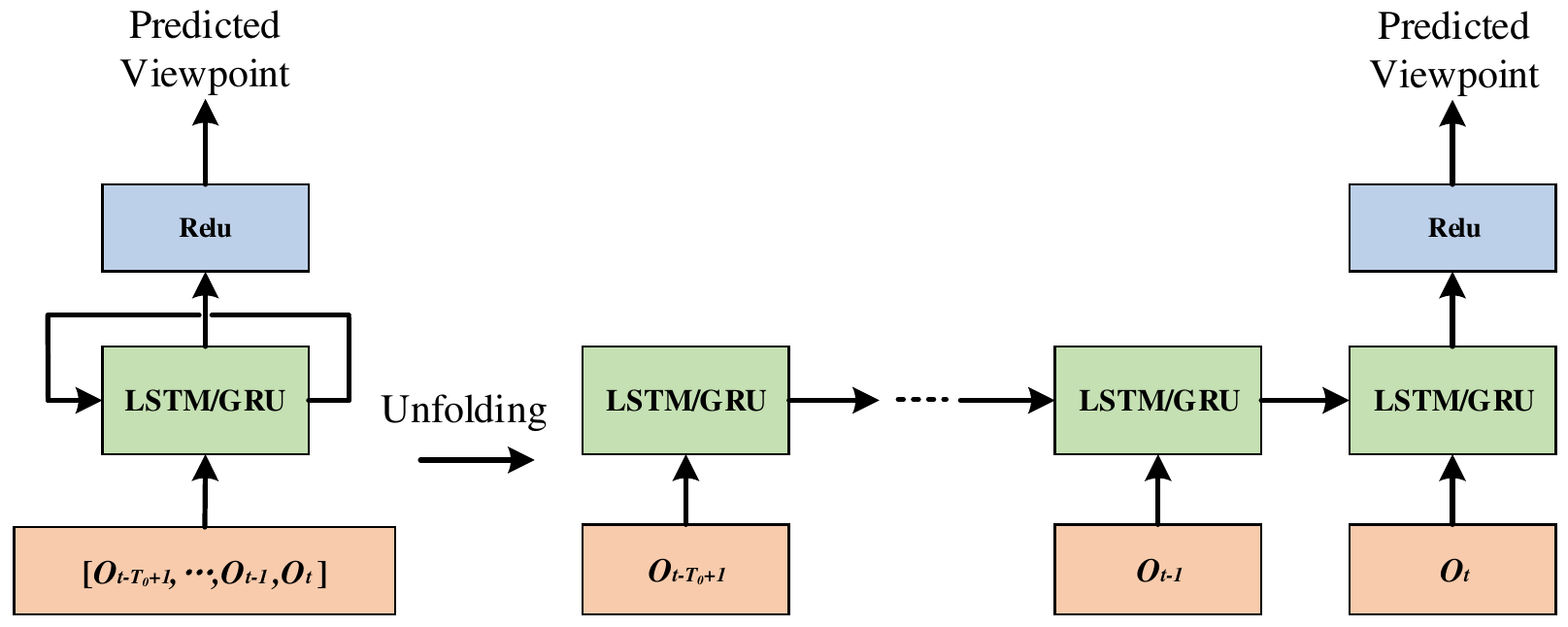}
    \caption{Proposed LSTM/GRU architecture (left) with its unfolding structure (right).}
    \label{basic_modules}
\end{figure}

\subsubsection{Long-short Term Memory/Gated Recurrent Unit}
To capture the dynamics in viewpoint of the VR user for the $(t+1)$th time slot, we use not only the most recent observation $O_t = \{O_t^1,O_t^2,...,O_t^{K}\}$, where $O_t^k = \{(X_t^k,Y_t^k,Z_t^k)\}$ is the actual viewpoint of the $k$th VR user at the $t$th time slot, but also the previous observations $H_t = \{O_{t-T_o+1},...,O_{t-2},O_{t-1}\}$, where $T_o$ is the size of the memory window. In order to recognize the viewpoint in continuous time slots, we leverage a RNN model with parameters $\bm{\theta}^{\rm{RNN}}=\{\bm{W}^{\rm{RNN}}, \bm{b}^{\rm{RNN}}\}$, where $\bm{W}^{\rm{RNN}}$ and $\bm{b}^{\rm{RNN}}$ are the neurons' weights and bias vector of the RNN. The RNN is capable of capturing time correlation of the viewpoint of the VR user, which can help learn the time-varying viewpoint for better prediction accuracy.

The LSTM/GRU layer contains multiple standard LSTM/GRU units and receives the current and historical observations $[O_{t-T_o+1},...,O_{t-1},O_{t}]$ at the $t$th time slot and is connected to an output layer with a Relu non-linearity activation function, which is shown in Fig. 13. The Relu layer outputs the predicted viewpoint of the VR user for the $(t+1)$th time slot. To update the model parameter $\bm{\theta}^{\rm{RNN}}$, we first use MSE to calculate the loss function, and then use the standard SGD via BackPropagation Through Time (BPTT) \cite{BPTT}. At the $(t+1)$th time slot, $\bm{\theta}^{\rm{RNN}}$ can be updated as
\begin{equation}
    \bm{\theta}_{t+1,l}^{\rm{RNN}} = \bm{\theta}_{t,l}^{\rm{RNN}}-\lambda^{\rm{RNN}}\Delta\mathcal{L}_{t,l}^{\rm{RNN}}(\bm{\theta}_{t,l}^{\rm{RNN}}),
\end{equation}
where $\lambda^{\rm{RNN}}\in(0,1]$ is the learning rate of the RNN, $\Delta\mathcal{L}_{t,l}^{\rm{RNN}}(\bm{\theta}_{t,l}^{\rm{RNN}})$ is the gradient of the loss function $\mathcal{L}_{t,l}^{\rm{RNN}}(\bm{\theta}_{t,l}^{\rm{RNN}})$ to train parameters of the RNN. $\mathcal{L}_{t,l}^{\rm{RNN}}(\bm{\theta}_{t,l}^{\rm{RNN}})$ can be computed by the MSE as
\begin{equation}
    \mathcal{L}_{t,l}^{\rm{RNN}}(\bm{\theta}_{t,l}^{\rm{RNN}}) = \|\bm{\phi}_{t,l}^{\rm{RNN}}-\hat{\bm{\phi}}_{t,l}^{\rm{RNN}}\|_2,
\end{equation}
where $\bm{\phi}_{t,l}^{\rm{RNN}}$ is the desired output of the $l$th layer in RNN, $\hat{\bm{\phi}}_{t,l}^{\rm{RNN}}$ is the dependence of the RNN's output to the $l$th layer's parameters. The proposed Proactive retransmission scheme integrated into the online RNN is presented in Algorithm 1.

\begin{algorithm}[!h]
\begin{algorithmic}[1]

\caption{The Proactive retransmission scheme integrated into Online Learning Algorithms with $n$-order LR, NN and LSTM/GRU}
\STATE Initialize the order $n$ of LR, parameters $\bm{\theta}^{\rm{LR}}$ or $\bm{\theta}^{\rm{NN}}$ or $\bm{\theta}^{\rm{RNN}}$ , and sliding window size $T_w$.
\STATE Use K Cross Validation to train the parameters of the $n$-order LR, NN and RNN learning model.

\FOR{t = 1,...,T}
    \STATE Get historical viewpoint from the $(t-T_w)$th time slot to the $(t-1)$th time slot from the updated sliding window.
    \STATE Use the updated online $n$-order LR, NN, LSTM/GRU to predict the viewpoint of the VR user for the $t$th time slot.
    \STATE The VR user transmits its actual viewpoint of the $t$th time slot via uplink transmission with the Proactive retransmission scheme.
    \IF{the uplink transmission is successful}
        \STATE Update parameters $\bm{\theta}_t^{\rm{LR}}$ or $\bm{\theta}_t^{\rm{NN}}$ or $\bm{\theta}_t^{\rm{RNN}}$ of the $n$-order LR, NN and RNN learning model via (12), (15) and (16).
        \STATE Update the sliding window with the actual required viewpoint of the $t$th time slot.  
    \ELSE 
        \STATE $\bm{\theta}_{t-1}^{\rm{LR}}\rightarrow\bm{\theta}_t^{\rm{LR}}$ or $\bm{\theta}_{t-1}^{\rm{NN}}\rightarrow\bm{\theta}_t^{\rm{NN}}$ or $\bm{\theta}_{t-1}^{\rm{RNN}}\rightarrow\bm{\theta}_t^{\rm{RNN}}$.
        \STATE Update the sliding window with null of the $t$th time slot.
    \ENDIF
\ENDFOR
\end{algorithmic}
\end{algorithm}

\section{Simulation Results}
In this section, we examine the effectiveness of our proposed offline and online learning algorithms on the upink viewpoint prediction of VR users under the Proactive retransmission scheme. We set the size of the sliding window as 10, and the size of the prediction window as 1. For the $n$-order LR, we consider $n=15$. For the NN, we use the fully-connected NN with two hidden layers, where the first and second layers have 12 and 10 neurons, respectively. For the RNN, it has one hidden layer with 12 units. The learning rate for learning algorithm is 0.001. For the uplink transmission, we set $M=30$, $\alpha=3$, $\rm{TTI}=0.125~\rm{ms}$, $R_{\rm{th}}^{\rm{up}}=2~ \rm{MB/s}$, $\sigma^2=-110~\rm{dBm}$, and $K_{\rm{re}}=8$. Consider a limited square area whose side length is 100 meters. For simplicity, we use ``w/ Proac" and ``w/o Proac" to represent ``with Proactive Retransmission" and ``without Proactive Retransmission'', respectively. Meanwhile, in the Genie-aided scheme, the online learning model is trained with the correct actual viewpoint of each VR user at each time slot, which is the upper bound of the online learning algorithm with proactive retransmission scheme and cannot be reached in the practical wireless VR system.

\subsection{VR Dataset Processing}
We first save all the VR user samples in a MATLAB file. Then, we use Python 3.6 to delete the useless rows and columns, and import the VR user data into training and testing datasets. According to \cite{Bao2}, the motion of the VR user has strong short-term auto-correlations in all three dimensions. Due to the fact that auto-correlations are much stronger than the correlation between these three dimensions, the angles in each direction can be trained independently and separately. According to Fig. 5, 6 and 7, we can obtain that the range of $Y$ angle distribution is much larger than that of $X$ and $Z$. Therefore, for simplicity, we use offline and online learning algorithms to predict $Y$ angle of VR users in this section, however, our algorithms can also be used for the prediction of $X$ and $Z$ angles.

\subsection{Viewpoint Prediction}

The simulation results of our proposed two viewpoint prediction learning models, namely, one training model for single VR video, and one training model for all VR videos, are introduced as follows:

(a) \textbf{One Training Model for One VR Video:} In this scheme, for each VR video, we use the VR user samples in the training datasets to train the offline and online learning models to predict the $Y$ angle of VR users in the testing datasets, and average the prediction error of all VR videos.

\begin{figure}[!h]
    \centering
    \includegraphics[width=3.5 in]{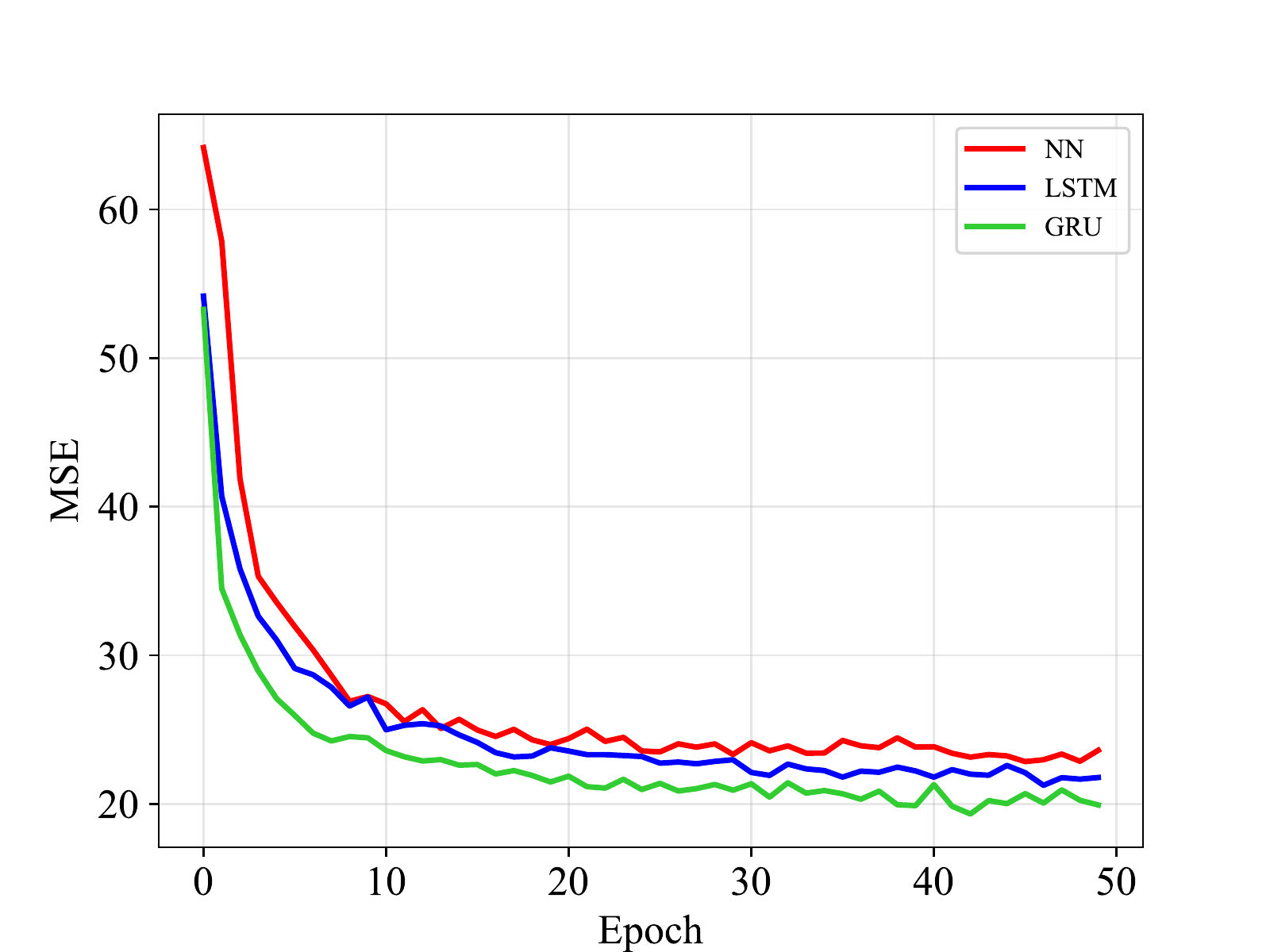}
    \caption{Loss of offline NN, LSTM and GRU algorithms of each epoch.}
    \label{basic_modules}
\end{figure}

Fig. 14 plots the loss of offline NN, LSTM and GRU algorithms of each epoch. It is seen that the performance of offline GRU algorithm outperforms that of LSTM and NN. This is because the structure of the LSTM is more complex than that of GRU, so that the parameters of the GRU can be trained faster and easier to be modified \cite{GRU_and_LSTM1}.

\begin{figure}[!h]
    \centering
    \includegraphics[width=3.5 in]{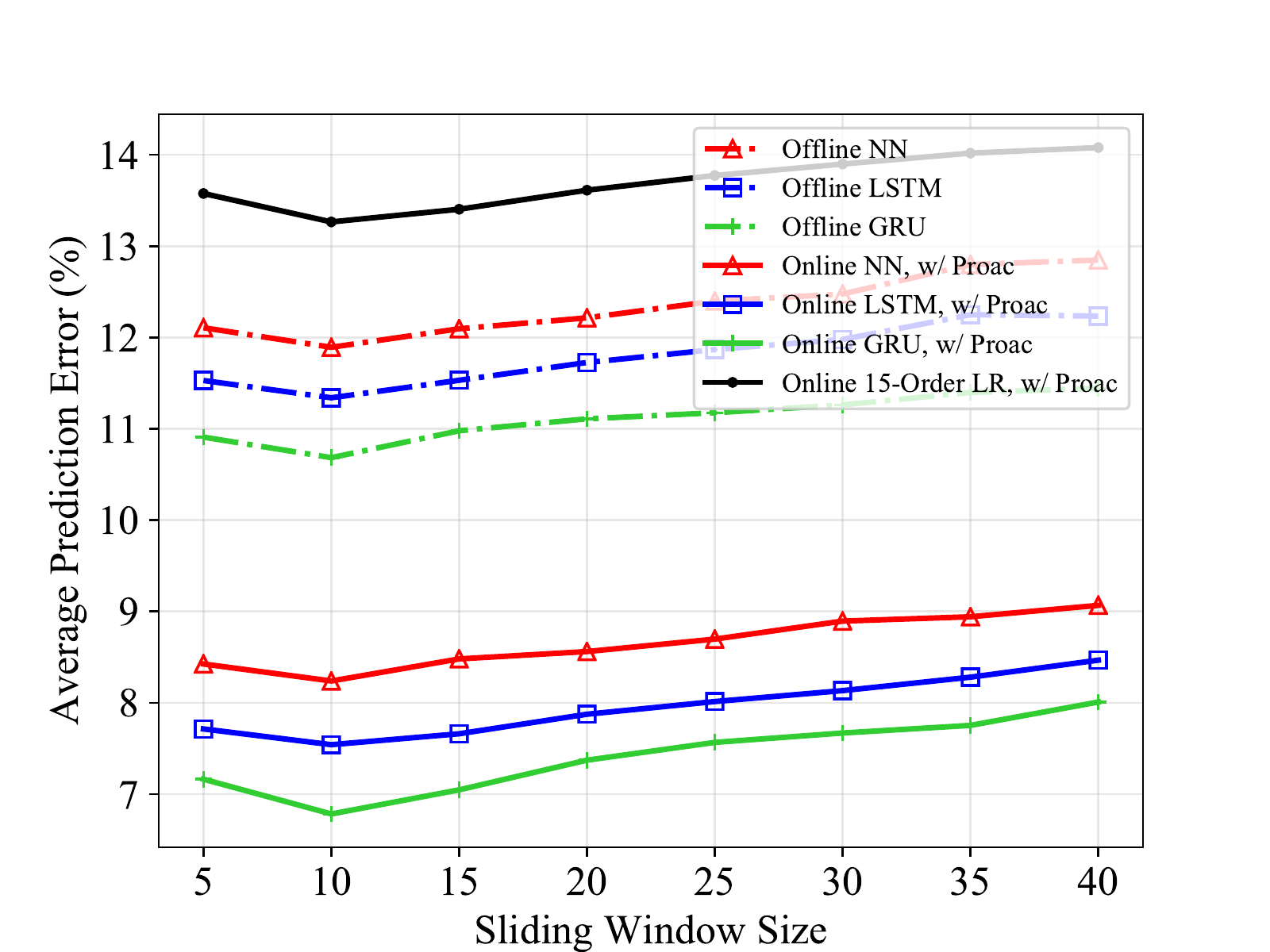}
    \caption{Average prediction error of offline/online learning algorithms via different size of sliding window with Proactive retransmission scheme.}
    \label{basic_modules}
\end{figure}

Fig. 15 shows the average prediction error of offline/online learning algorithms via different size of sliding window for uplink VR viewpoint transmission with proactive retransmission scheme. It is noted that the average prediction error of the offline/online 15-order LR, NN, LSTM and GRU algorithms is not significantly affected by changing the size of sliding window due to their capability to adapt to the viewpoint preference. When the size of the sliding window is 10, it can obtain the best performance.

\begin{figure}[!t]
    \centering
    \includegraphics[width=3.5 in]{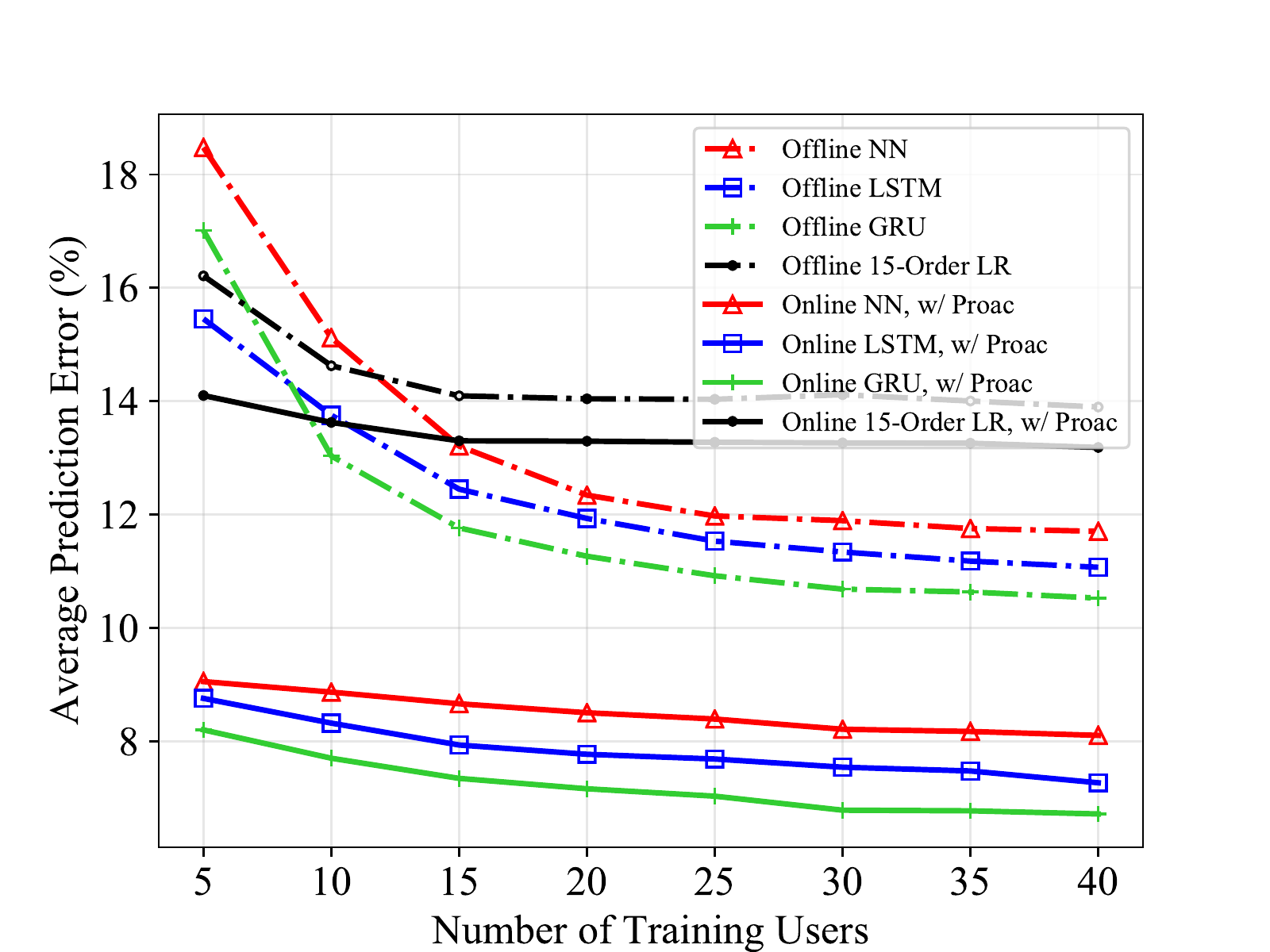}
    \caption{Average prediction error of different number of VR users in the training dataset to train the learning model via offline/online $15$-order LR, NN, LSTM and GRU with Proactive retransmission scheme.}
    \label{basic_modules}
\end{figure}

Fig. 16 plots the average prediction error of different number of VR users in the training dataset to train the learning model via offline/online 15-order LR, NN, LSTM and GRU for uplink VR viewpoint transmission with proactive retransmission scheme. For the offline learning algorithms, we observe that the average prediction error becomes smaller with increasing number of VR users. With increasing number of VR users, the offline learning algorithms can be trained to adapt to the viewpoint of the VR users much more accurately. It is also seen that the performance of the LSTM/GRU is better than that of the NN. This is because the LSTM/GRU is able to capture the correlation of the viewpoint in continuous time slots. In addition, it can be seen that the average prediction error of $15$-order LR algorithm is much higher than that of offline/online NN, LSTM and GRU. It is because the learning structure of the LR algorithm is simplier than that of NN, LSTM and GRU, and its ability to be fit for the nonlinear VR data is worse than that of the NN, LSTM and GRU. In addition, LR algorithm may get overfit with so many VR users training the LR model.

Meanwhile, for the proactive retransmission scheme integrated into the online learning algorithm in Fig. 16, it is interesting to note that its average prediction error is much smaller than that of offline learning algorithms and changes slightly with the increasing number of VR users. This is due to that through updating the parameters in the trained learning model, the online learning algorithm is able to adapt to the viewpoint preference of new VR users over time. Thus, the prediction accuracy can be improved.

\begin{figure}[!t]
    \centering
    \includegraphics[width=3.5 in]{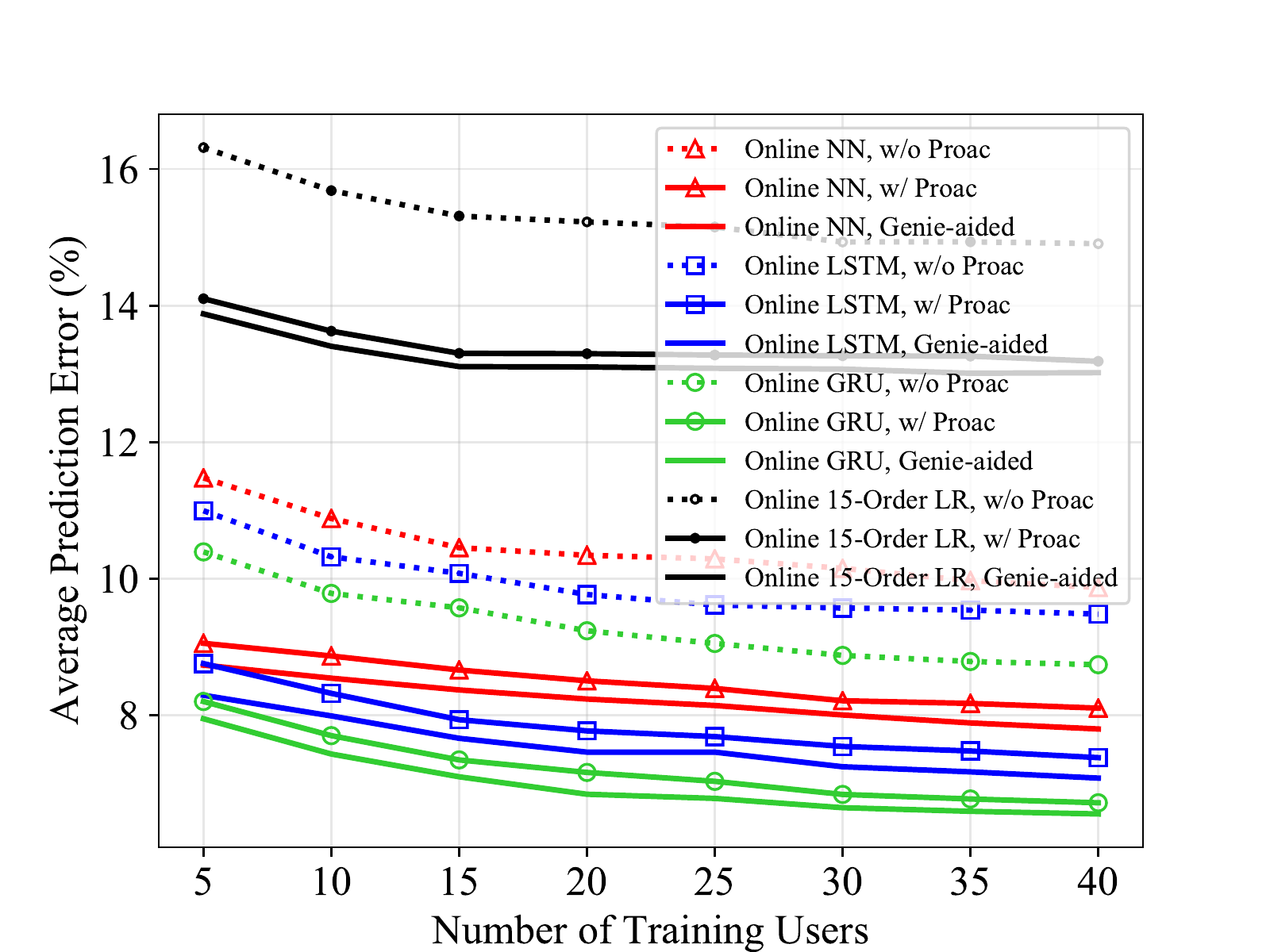}
    \caption{Average prediction error of different number of VR users in the training dataset to train the learning model via online $15$-order LR, NN, LSTM and GRU with/without Proactive retransmission scheme.}
    \label{basic_modules}
\end{figure}

Fig. 17 plots the average prediction error for various number of VR users in the training dataset to train the learning model via online $15$-order LR, NN, LSTM and GRU for uplink VR viewpoint transmission with/without proactive retransmission scheme. We can observe that the performance of the proactive retransmission scheme integrated into the online learning algorithm is better than that without proactive retransmission scheme and is close to the performance of the Genie-aided scheme. In the uplink viewpoint transmission without proactive retransmission, each VR user only transmits its actual viewpoint to the SBS once even this transmission fails. This transmission failure is usually because of the unstable channel state and the interference from other VR users. To cope with this, the proactive retransmission scheme is applied here to improve the success transmission of uplink transmission \cite{yanliu}, and the online learning algorithms are capable of better capturing historical trends of viewpoint preference of the VR user, which can further improve the prediction accuracy. While in the Genie-aided scheme, the uplink transmission at each time slot for each VR user is assumed to be successful.

(b) \textbf{One Training Model for All VR Videos:} In this model, for all 16 VR videos, we consider 4 Cross Validation shown in Fig. 11 and use the VR user samples in the training datasets to train offline/online $15$-order LR, NN and GRU learning models to predict $Y$ angle of VR users in the testing datasets.

\begin{figure}[!h]
    \centering
    \includegraphics[width=3.5 in]{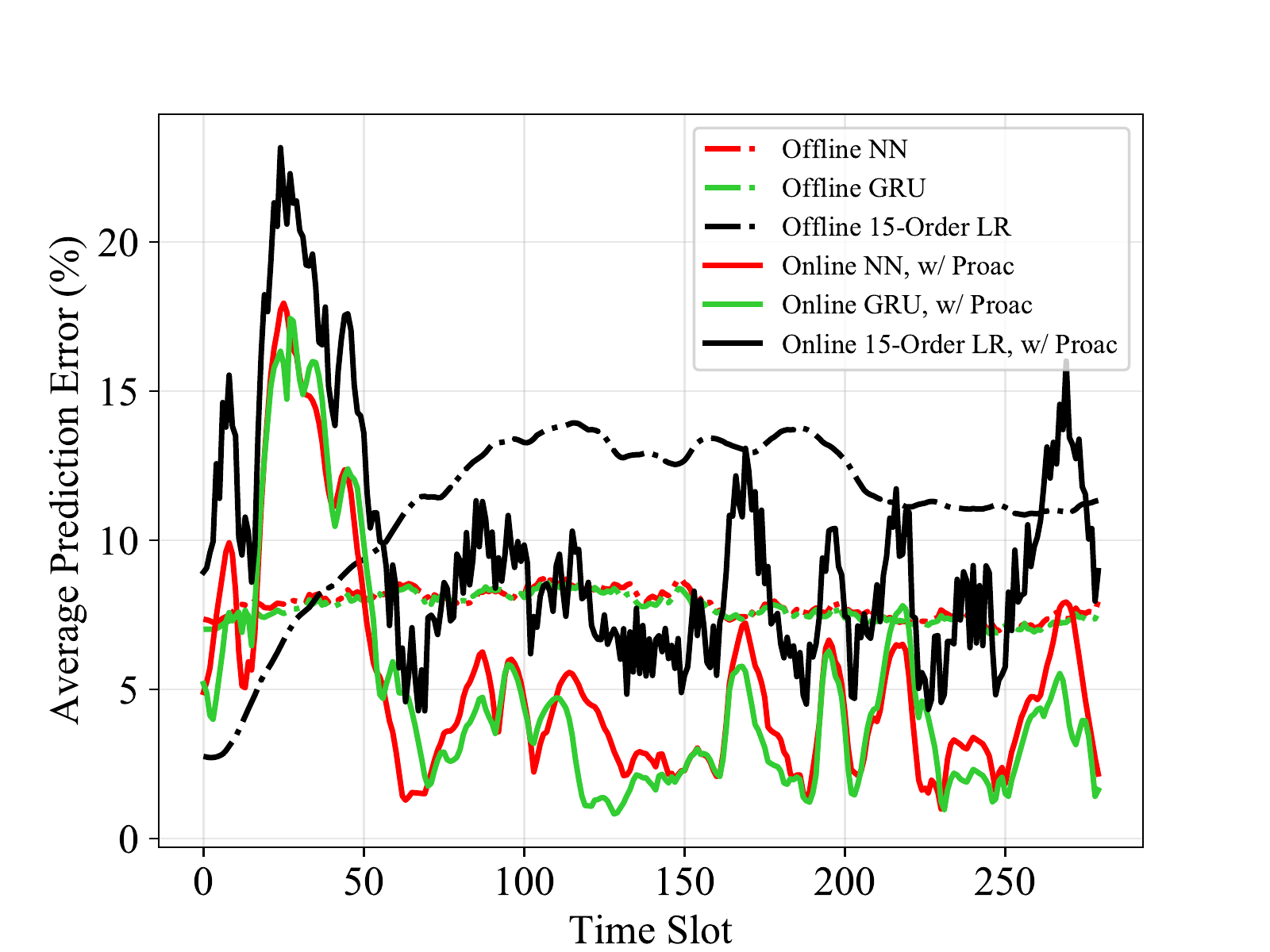}
    \caption{Average prediction error of offline/online $15$-order LR, NN and GRU with Proactive retransmission scheme in continuous time slots.}
    \label{basic_modules}
\end{figure}

Fig. 18 plots average prediction error of offline/online $15$-order LR, NN and GRU integrated with proactive retransmission scheme over continuous time slots. For the offline learning algorithms, it can be seen that the performance of the GRU is a bit better than that of the NN. Meanwhile, it can be observed that at the beginning 30 time slots, the performance of the $15$-order LR is better than that of NN and GRU. It is because according to Fig. 6, at the beginning, when the VR user watches the VR video, its viewpoint mainly focus on the zero point and the $15$-order LR fits well at the beginning. However, after 50 time slots, the performance of GRU is much better than that of $15$-order LR. This is due to that after 50 time slots, the viewpoint of the VR user will change substantially as shown in Fig. 6, and the GRU is able to capture the correlation of the viewpoint of the VR user over continuous time slots. 

Furthermore, it is also noted that at the beginning, there are large fluctuations in the performance of the proactive retransmission scheme integrated into the online learning algorithms. It is because the parameters in the online learning algorithms should be modified to capture the viewpoint preference of the VR user. In addition, when the viewpoint of the VR user changes over time, the online learning algorithms need to further update their parameters to be fit for the viewpoint changing of the VR users. Therefore, there are small fluctuations in the performance of the online learning algorithms.

\begin{figure}[!h]
    \centering
    \includegraphics[width=3.5 in]{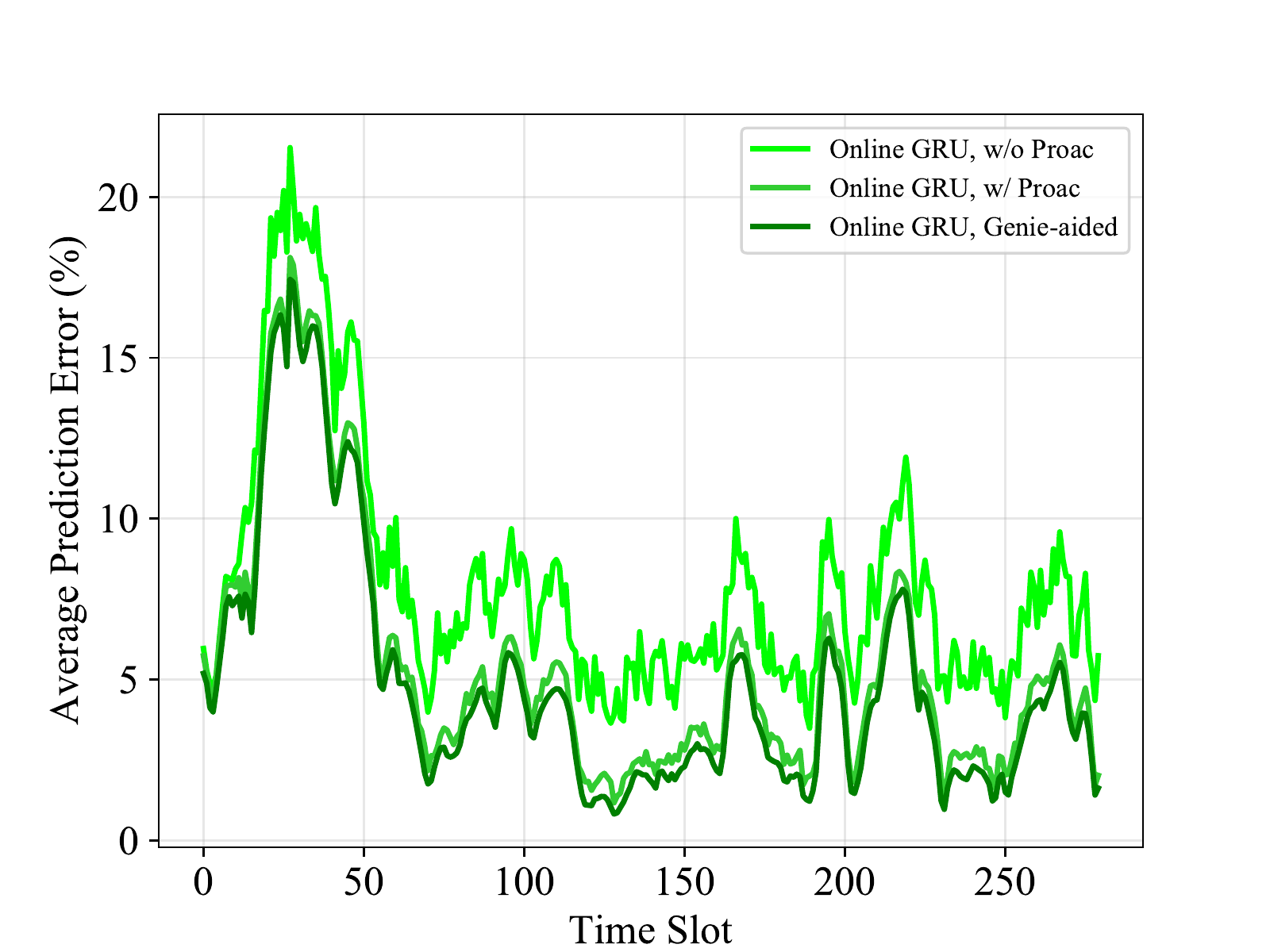}
    \caption{Average prediction error of online GRU algorithms with/without Proactive retransmission scheme in continuous time slots.}
    \label{basic_modules}
\end{figure}

Fig. 19 plots the average prediction error of online GRU algorithms of uplink viewpoint transmission with/without proactive retransmission scheme in continuous time slots. We can obtain that the performance of the proactive retransmission scheme with the online GRU algorithm is still better than that of the scheme without proactive retransmission scheme. Meanwhile, it can be seen that the performance of the Genie-aided online GRU algorithm slightly outperforms that of the online GRU algorithm with the proactive retransmission scheme, while their gap is small.

\section{Conclusions}
In this paper, offline and online learning algorithms for uplink wireless VR network with proactive retransmission scheme were developed to predict viewpoint of wireless VR users with real VR dataset. Specifically, for the offline learning algorithm, K Cross Validation was used to train offline $n$-order LR, NN and LSTM/GRU learning algorithms for each VR video and all VR videos. The trained offline learning algorithms were used to directly predict the viewpoint of the VR user. In the online learning algorithms, the online $n$-order LR, NN and LSTM/GRU algorithms would update their parameters according to the actual viewpoints delivered from the new VR users through uplink transmission, which could further improve the prediction accuracy. Meanwhile, a proactive retransmission scheme was introduced to the online learning algorithms to guarantee the reliability of uplink transmission. Simulation results shown that our proposed online GRU algorithm with the proactive retransmission scheme can achieve the highest prediction accuracy. Meanwhile, the single training model for each VR video, and for all VR videos achieved similar prediction accuracy.

%





\ifCLASSOPTIONcaptionsoff
  \newpage
\fi





%
%


\end{document}